\documentclass[12pt,preprint]{aastex}

\newcommand{\msun}{$M_{\sun}$}

\begin{document}
\title{Photometry of the Globular Cluster NGC 5466: Red Giants and Blue Stragglers}
\author{Nassissie Fekadu and Eric L. Sandquist\altaffilmark{1,2}}
\affil{Department of Astronomy, San Diego state University, 5500 Campanile
Drive, San Diego, CA 92182} \email{fekadu@sciences.sdsu.edu,
erics@sciences.sdsu.edu}
\author{Michael Bolte\altaffilmark{1}} \affil{University of California Observatories,
University of California, Santa Cruz, CA 95064}
\email{bolte@ucolick.org}
\altaffiltext{1}{Visiting Astronomer, Kitt Peak National Observatory, National
  Optical Astronomy Observatory, which is operated by the Association of
  Universities for Research in Astronomy, Inc. (AURA) under cooperative
  agreement with the National Science Foundation.}
\altaffiltext{2}{Guest User, Canadian Astronomy Data Centre, which is operated
  by the Dominion Astrophysical Observatory for the National Research Council
  of Canada's Herzberg Institute of Astrophysics.}

\begin{abstract}
  We present wide-field $BVI$ photometry
  for about 11,500 stars in the low-metallicity cluster NGC 5466. We
  have detected the red giant branch bump for the first time,
  although it is at least 0.2 mag fainter than expected relative to
  the turnoff. The number of red giants (relative to main sequence turnoff stars) is in excellent
  agreement with stellar models from the Yonsei-Yale and Teramo
  groups, and slightly high compared to Victoria-Regina models.
  This adds to evidence that an abnormally large ratio of red
  giant to main-sequence stars is not correlated with cluster metallicity.  We
  discuss theoretical predictions from different
  research groups and find that the inclusion or exclusion of helium diffusion
  and strong limit Coulomb interactions may be partly responsible.

We also examine indicators of dynamical history: the
mass function exponent and the blue straggler frequency. NGC 5466 has a
very shallow mass function, consistent with large mass loss and
recently-discovered tidal tails.  The blue straggler sample is significantly
more centrally concentrated than the HB or RGB stars.  We see no evidence of
an upturn in the blue straggler frequency at large distances from the center.
Dynamical friction timescales indicate that the stragglers should be {\it
  more} concentrated if the cluster's present density structure has existed
for most of its history. NGC 5466 also has an unusually low central density
compared to clusters of similar luminosity. In spite of this, the
specific frequency of blue stragglers that puts it right on the
frequency -- cluster $M_V$ relation observed for other clusters.

\end{abstract}

\keywords{equation of state --- globular clusters: individual (NGC 5466) ---
   stars: blue stragglers --- stars: evolution --- stars: luminosity function}

\section{Introduction}

The luminosity function (LF) is an observational tool used for
analyzing the post-main sequence evolutionary phases of low-mass
($\approx$ 0.5-0.8{\msun}) metal-poor stars in Galactic globular
clusters (GGC).  Because of their age and richness, GGC
typically contain hundreds of stars that have evolved off the
main sequence. The numbers of stars in evolved phases are directly
related to the evolutionary timescales and fuel consumed in each phase
\citep{ren88}, so that they present us with an opportunity to test
this aspect of stellar evolution models.
 
The results of the most stringent tests have been mixed. Repeated
studies of the metal-poor cluster M30 \citep{bolte,berg,guh98,san99}
have found an excess number of red giant branch (RGB) stars relative
to main sequence (MS) stars. \citet{stet91a} also uncovered an
apparent excess of stars in a combined LF of the metal-poor clusters
M68, NGC 6397, and M92. However, the LFs of more metal-rich clusters
show no discrepancy (M5: Sandquist et al. 1996; M3: Rood et
al. 1999; M12: Hargis, Sandquist, \& Bolte 2004). In a survey of 18
clusters, \citet{zoc00} found good agreement with model predictions
with the possible exception of clusters at the high metallicity end.

In this paper we present $BVI$ photometry of NGC 5466, a high galactic
latitude globular cluster ($l = 42.2\degr$ and $b=73.6\degr$), located
in the constellation of Bo\"{o}tes ($\alpha = 14^{h}05^m27\fs4$,
$\delta = +28\degr32\arcmin04\arcsec$ at a distance of R=15.9 kpc;
Harris 1996).  NGC 5466 is a loose cluster ($r_c = 1\farcm64$) with
extremely low metallicity ([Fe/H]$ =-2.22$) and subject to little or
no reddening, $(E(B-V)
\simeq 0)$ \citep{har96}.

In \S 2, we describe the process leading to the calibrated photometry, and
compare with previous studies of the cluster. In \S 3, we compare the observed
color-magnitude diagram and observed luminosity function with theoretical
models, focusing on the relative number of stars on the lower RGB and around
the MS turnoff. Finally, in \S \ref{bss}, we present a new examination of the
blue straggler population of NGC 5466.

\section{Observations and Data Reduction}

The data used in this study were obtained with the Kitt Peak National
Observatory (KPNO) 0.9 m telescope ($0\farcs68$ pix$^{-1}$) on the nights of
UT dates 1995 May 4, May 5, and May 9. A complete list of the image frames, 
exposure times, and observing conditions is given in Table \ref{tab31}.  

The images obtained on the three nights were processed using
IRAF\footnote{IRAF(Image Reduction and Analysis Facility) is distributed by
  the National Optical Astronomy Observatories, which are operated by the
  Association of Universities for Research in Astronomy, Inc., under contract
  with the National Science Foundation.} tasks and packages. The 
reduction involved subtraction and trimming of the overscan region of
all images, subtraction of a master bias frame from flats and object frames,
and flat fielding of the object frames using images taken at twilight.
Profile-fitting photometry was done using the DAOPHOTII/ALLSTAR programs
\citep{stet87}.

We also reduced archival ground-based photometry of the cluster core taken
with the High-Resolution Camera (HRCam)
on the 3.6 m Canada-France-Hawaii Telescope (CFHT).  The $V$
and $I$ images were taken 30 and 31 May 1992 (observers J. Heasley and C.
Christian), and have not previously been described in the literature. The CCD
images had 1024 $\times$ 1024 pixels, 0\farcs13 per pixel, and excellent
seeing (0.4 - 0.5 arcsec). The images were reduced using the archived bias and
twilight flat frames, and following a procedure similar to that for the KPNO
data.  This allowed us to get excellent photometry to 2 magnitudes below
the turnoff in the cluster core. These images were used entirely for blue
straggler identification (see \S \ref{bss}).

\subsection{Calibration against Primary Standard Stars}

The conditions at KPNO on 1995 May 9 were photometric, and
\citeauthor{lan92} standard star fields were observed at a range of air
masses to determine photometric transformation coefficients. The standard
values used for the calibration were chosen from the large compilation
of \citet{stet00}, which is set to be on the same photometric scale as
the earlier \citet{lan92} values.

\begin{figure}
\plotone{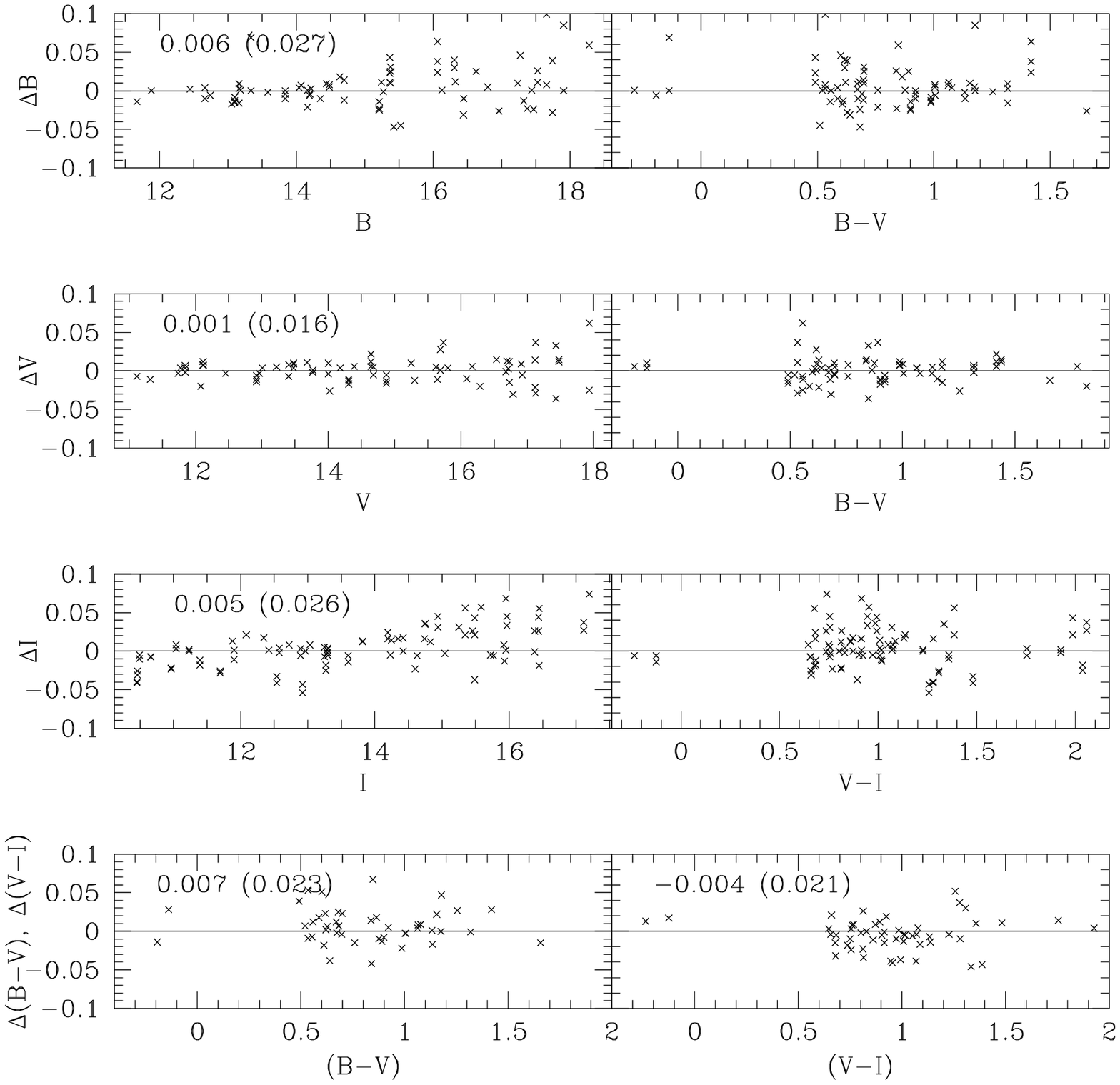}
\caption{Photometric residuals (in the sense of this study minus
those of \citealt{lan92} and \citealt{stet00}) of primary
standard stars. The median residuals are listed in the panels with the
semi-interquartile range (half the magnitude difference between the 25\% and
75\% points in the ordered list of residuals) given in parentheses. 
\label{fig311}}
\end{figure}

We conducted photometry on the standard stars and isolated cluster
stars using multiple synthetic apertures. We then used the DAOGROW
\citep{stet90} program to construct growth curves to
extrapolate measurements to a common aperture size. Using the CCDSTD
program, the standard star transformation equations were found to be:
\[ b = B + a_{o}+(-0.069\pm0.005)(B-V)+(0.255\pm0.014)(X-1.0)\]
\[ v = V + b_{o}+(0.027\pm0.004)(B-V) +(0.172\pm0.010)(X-1.0)\]
\[ i = I + c_{o}+(-0.013\pm0.005)(V-I) +(0.145\pm0.018)(X-1.0) \]
where $X$ is the airmass, $v$, $b$ and $i$ are instrumental magnitudes, and
$V$, $B$, and $I$ are standard magnitudes. These calibration equations are
different than those used for our analysis of M10 \citet{pol05} and M12
\citet{harg04}, which were observed on the same night, because our $I-$band
exposures did not go as deep as the $B$ and $V$ exposures. As a result, the
$(B-V)$ color was a better choice for calibrating the $V$ photometry down to
our faintest observed stars. The calibrated measurements for the standard
stars are compared with catalogue values in Fig. \ref{fig311}.

We note that there is slight evidence of trend in the residuals for the $I$
band versus magnitude, which might indicate nonlinearity. This impression is
caused by one observation of the PG1323-086 field. We did, however, have an
additional observation of the same field on the same night having the same
exposure time that does not show the same (small) trend. Because we do not
have any reason to eliminate the frame and because its elimination has a
minimal effect on the transformation coefficients, we have decided to retain
the measurements from the image.

\subsection{Calibration against Secondary Standard Stars} 

Aperture photometry for 165 cluster stars was used to calibrate the
point-spread function (PSF) photometry for the cluster. These secondary
standard stars were chosen based on relatively low measurement errors and
location in relatively uncrowded regions of the cluster. They were
chosen from the asymptotic giant branch (AGB), upper RGB, and horizontal
branch (HB) of the cluster in order to cover the entire range of colors
covered by cluster stars.

\begin{figure}
\plotone{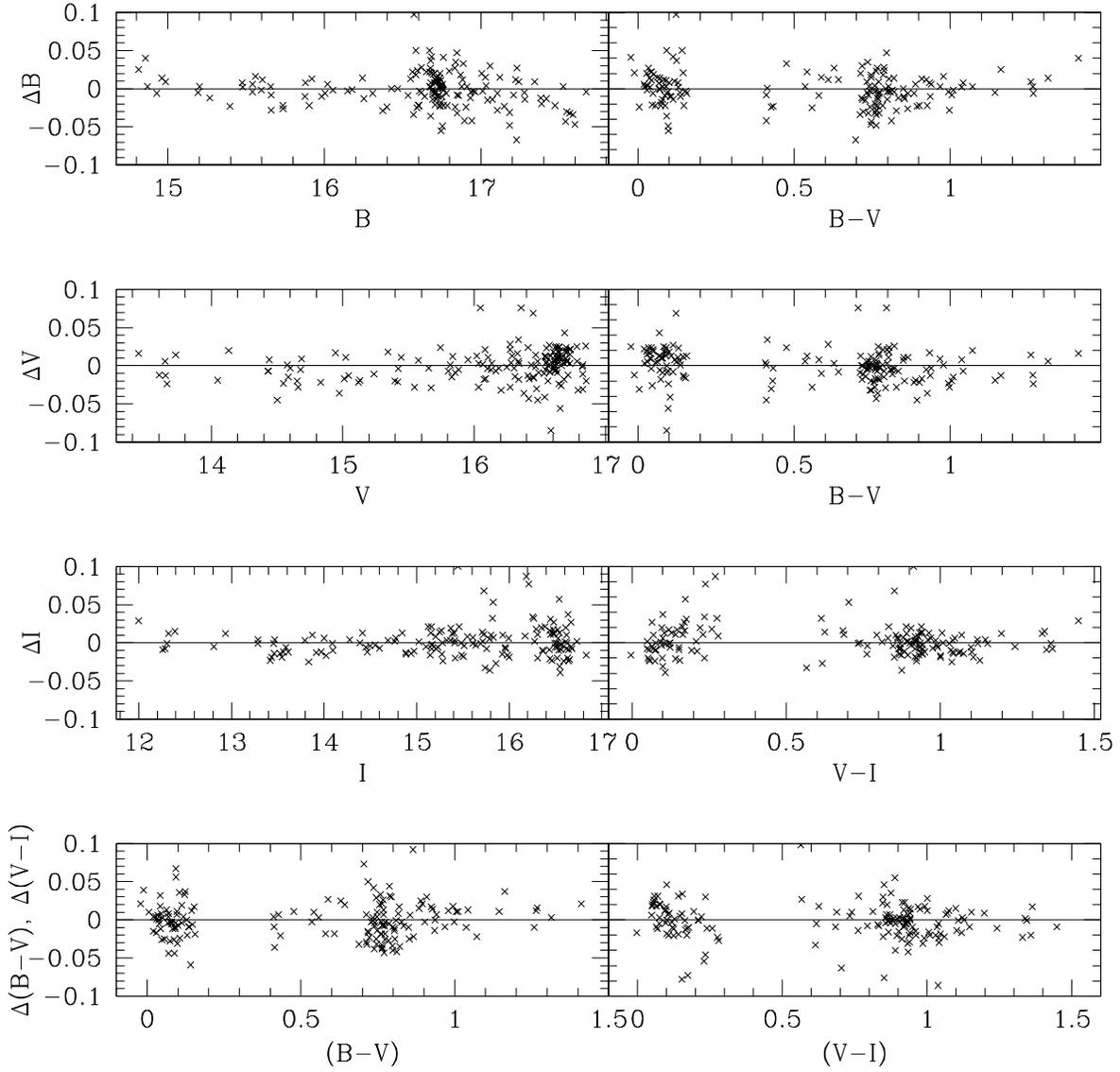}
\caption{Photometric residuals (in the sense of the final PSF photometry 
  minus standard aperture photometry values) of secondary standard stars.
\label{fig312}}
\end{figure}

The PSF-fitting photometry for the three nights of data was combined and
averaged after zero-point differences among the frames had been determined and
corrected. The zero-point corrections to the standard system were determined
after fixing the color-dependent terms at the values measured in the primary
standard star calibration. (This was also done in our studies of M10 and M12.)
In Fig. \ref{fig312}, it can be seen that this procedure does not introduce
systematic color- or magnitude-dependent errors.

\subsection{Comparison with Previous Studies} 

\begin{figure}
\plotone{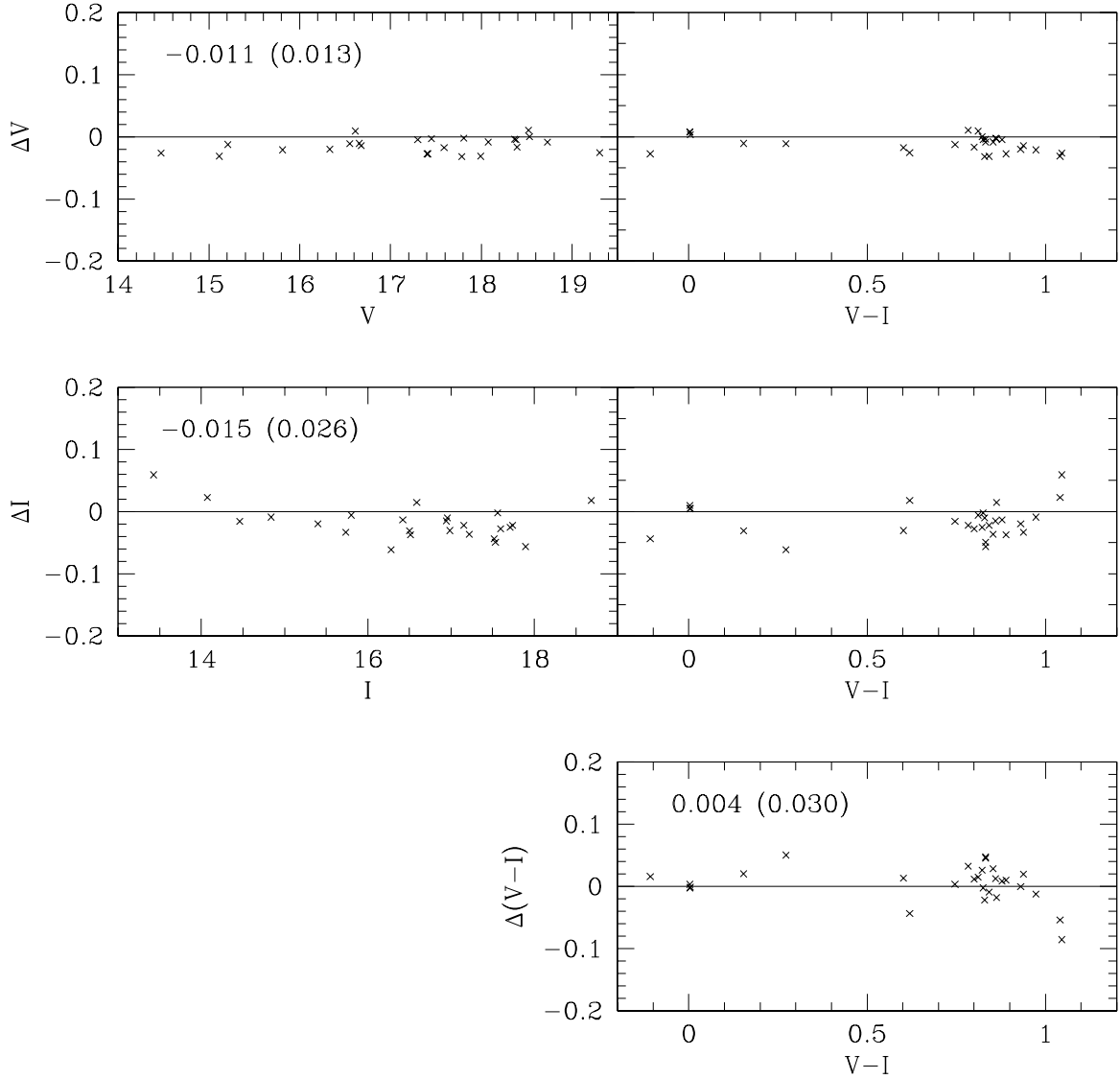}
\caption{Residuals (in the sense of this study minus \citealt{stet00}) from 
  the star-by-star comparison. The median residuals are listed in the panels
  with the semi-interquartile range (see Fig.\ref{fig311}) given in
  parentheses.
\label{fig315}}
\end{figure}

\begin{figure}
\plotone{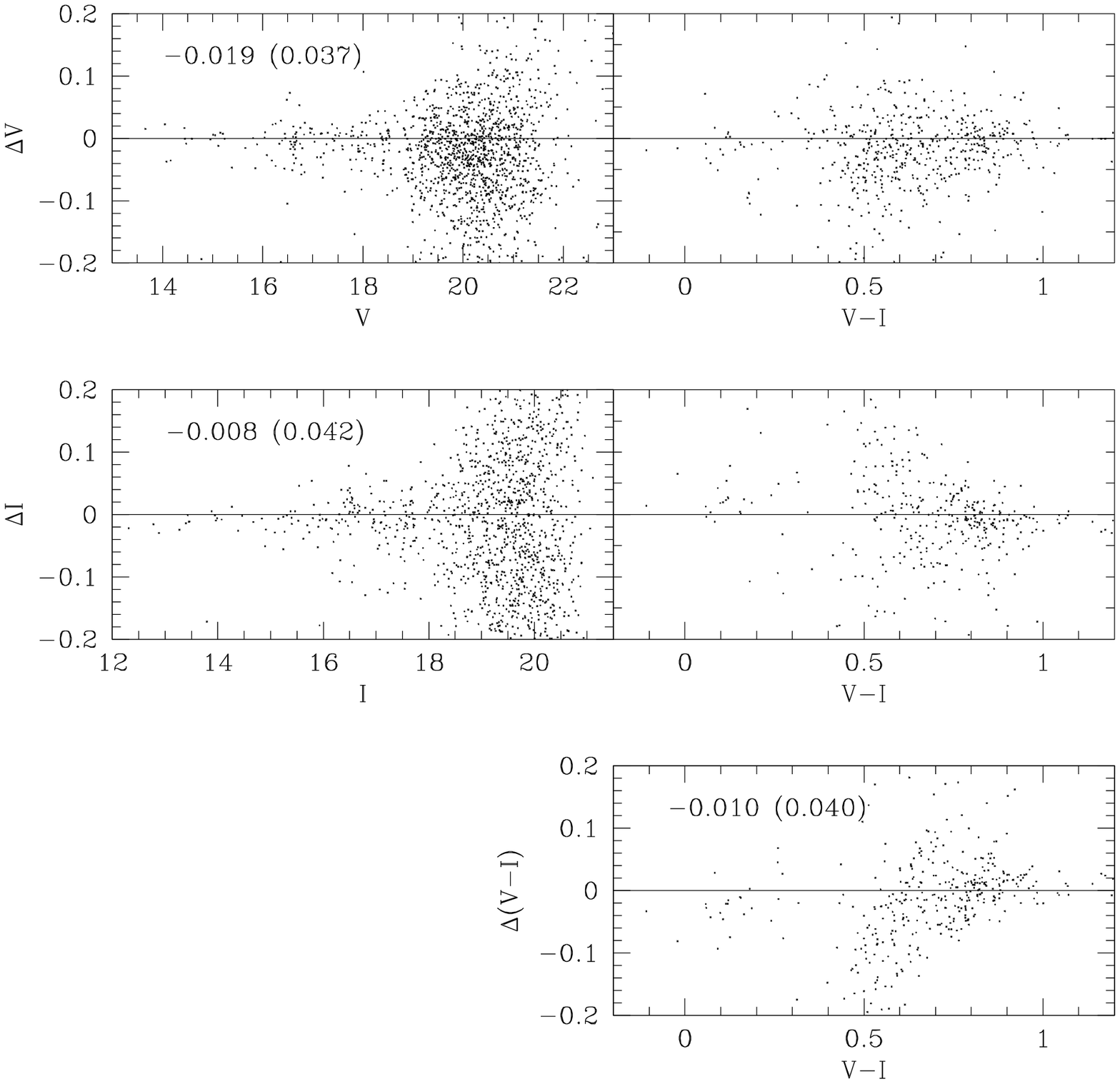}
\caption{Residuals (in the sense of this study minus \citealt{ros00}) from 
  the star-by-star comparison. The median residuals and the plots versus color
  have been restricted to brighter stars ($V < 20$ and $I < 19$) to make the
  comparisons clearer. The numbers in parentheses are the semi-interquartile
  ranges (see Fig.\ref{fig311}).
\label{fig314}}
\end{figure}

\begin{figure}
\plotone{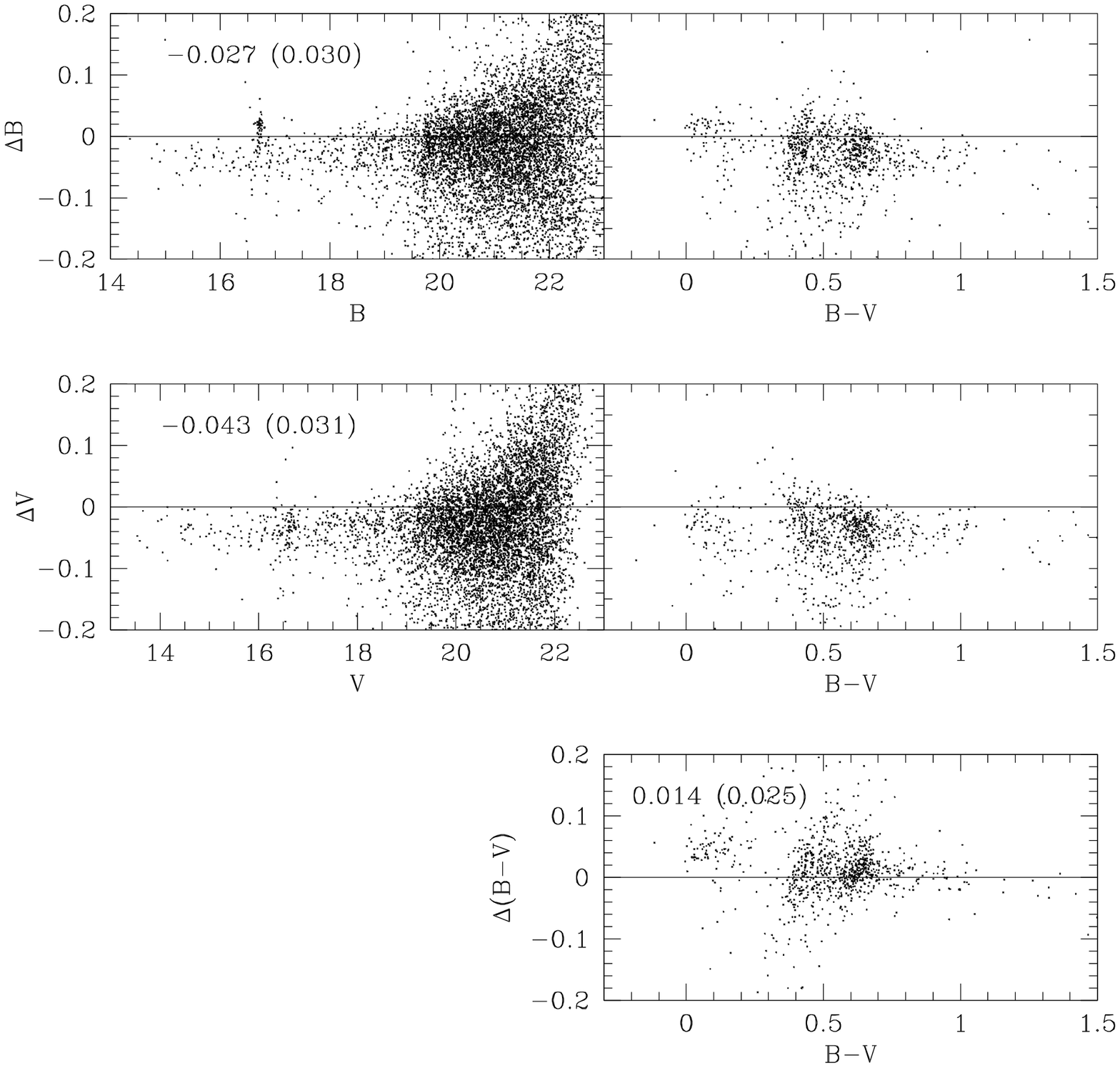}
\caption{Residuals (in the sense of this
  study minus \citealt{jeo04}) from the star-by-star comparison. The median
  residuals and the plots versus color have been restricted to brighter stars
  ($B < 20$ and $V < 19.5$) to make the
  comparisons clearer. The numbers in parentheses are the semi-interquartile
 ranges (see
  Fig.\ref{fig311}).
\label{fig313}}
\end{figure}

We compared our photometric data set to those of \citet{jeo04}, \citet{ros00},
and \citet{stet00}.  The magnitude and color comparisons ($BVI$ in this study
versus $VI$ in \citeauthor{stet00} and \citeauthor{ros00}, and $BV$ in
\citeauthor{jeo04}) as function of magnitude and color are shown in Figs.
\ref{fig315} -- \ref{fig313}. Though our calibrated magnitudes are slightly
brighter than those of \citet{stet00}, the differences are small, and there is
no color trend. The offsets compared to the \citet{ros00} are larger, but
again there are no clear color trends. The offsets compared to the
\citet{jeo04} data are also significant, but more notable are slight trends 
with color.

\subsection{Calculation of the Luminosity Function}

Artificial star tests were performed to empirically measure the precision of
our photometry and to correct for incompleteness in the detection of stars.
We followed the procedure described by
\citet{harg04} for the calculation of incompleteness corrections as function of
position and magnitude. 

The inputs used for producing the artificial star tests were the reduced $B$
and $V$ CCD frames, PSFs for each object frame, fiducial lines, and an
estimate for the initial LF \citep{san96}. Artificial stars were randomly
placed in cells on a spatial grid and the entire grid was then shifted
randomly from run to run in order to ensure the whole imaged field was tested
\citep{pio99}.  Each star was placed in a consistent position relative to the
cluster center on each image. If a detected star was found to coincide with
the input position of an artificial star, it was added to the archive. The new
images were reduced using the same procedure applied to the original data set.
In this study, a total of 100,000 artificial stars from 50 separate runs were
added. The number of artificial stars per trial was chosen so that the
effects of crowding on the photometry was qualitatively unchanged.

\begin{figure}
\plotone{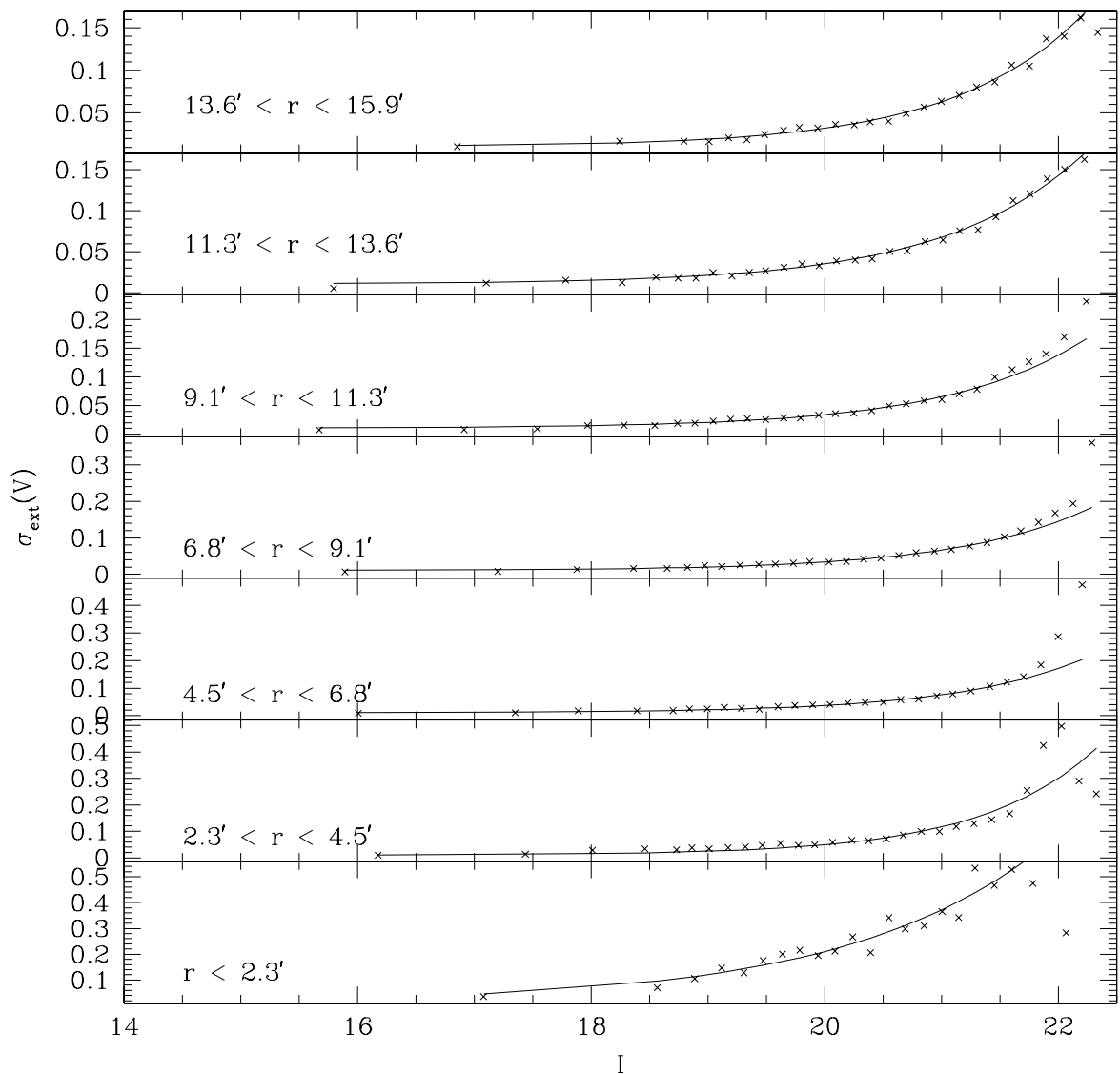}
\caption{External $V$ magnitude errors $\sigma_{ext}(V)$ as a function of
  radius and magnitude determined from artificial star tests, with exponential
  fits shown by the solid lines.
\label{fig613}}
\end{figure}

\begin{figure}
\plotone{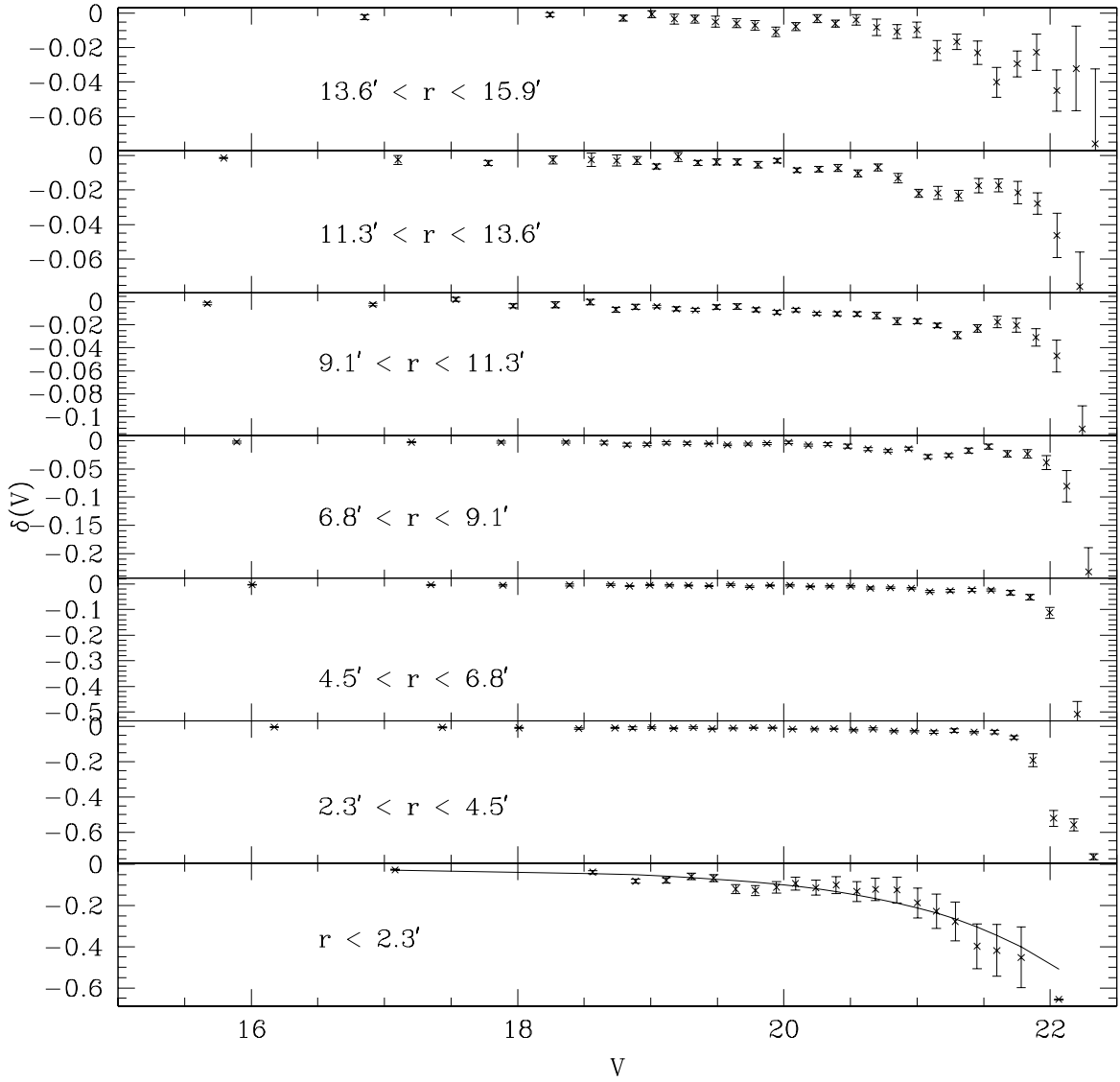}
\caption{Magnitude biases $\delta(V)$ determined from artificial star 
  tests as a function of radius and magnitude.
\label{fig614}}
\end{figure}

\begin{figure}
\plotone{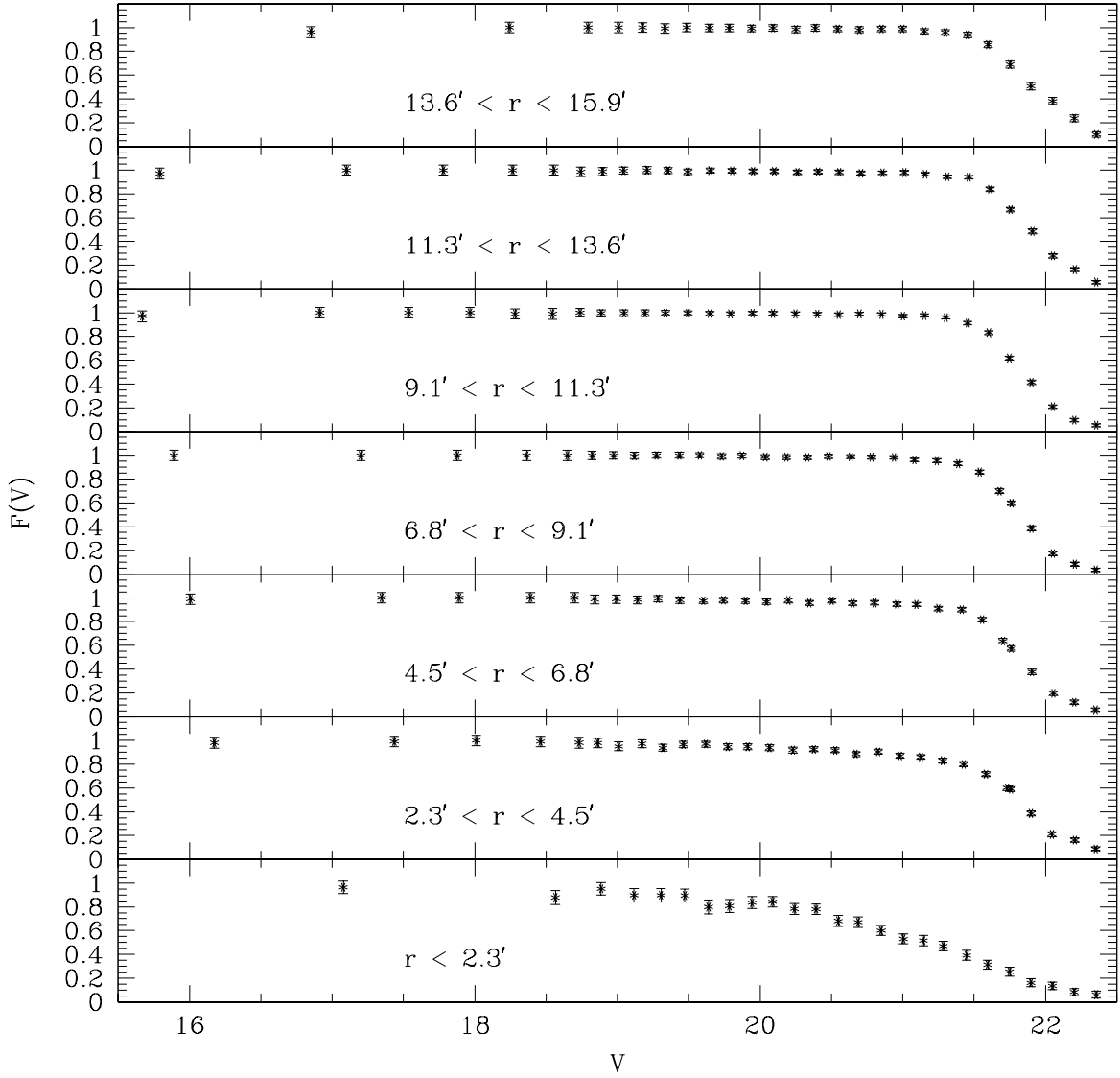}
\caption{Total recovery probability $F(V)$ determined from artificial star 
  tests as a function of radius and magnitude.
\label{fig615}}
\end{figure}

The recovered artificial stars were used to calculate 1) median magnitude and
color biases ($\delta_{V}$ and $\delta_{B-V}$, where
$\delta=\mbox{median}[output-input]$), 2) median external error estimates
($\sigma_{ext}(V)= \mbox{median}[\delta_{V}-\mbox{median}(\delta_{V})]/0.6745$
and $\sigma_{ext}(B-V)$), and 3) total recovery probabilities ($F(V)$, which
is the fraction of the stars that were recovered with any output magnitude) in
bins according to projected radius and magnitude. The values for the above
quantities are plotted in Figs.\ref{fig613} -- \ref{fig615}.

\begin{figure}
\plotone{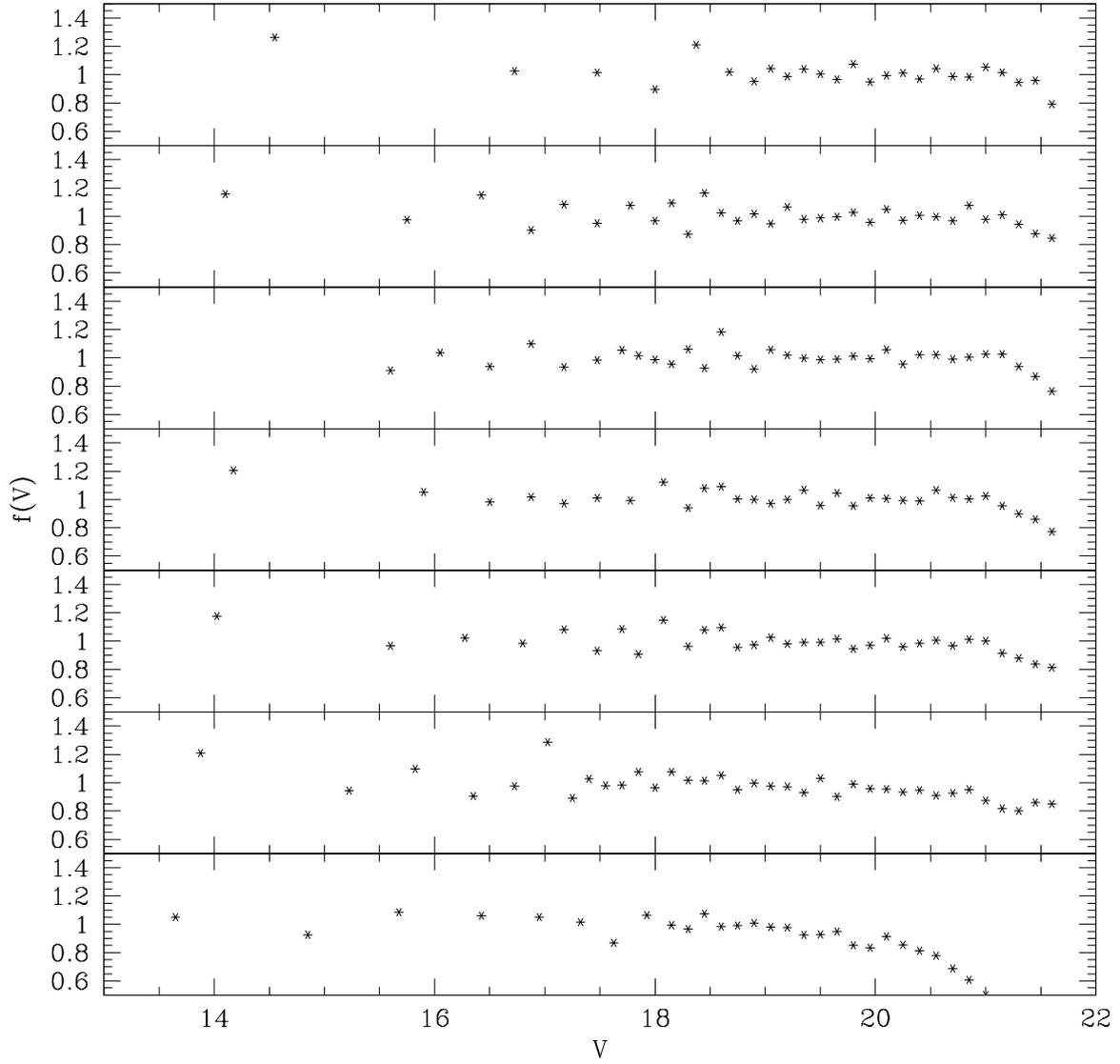}
\caption{Completeness fraction $f(V)$ determined from artificial star 
  tests as a function of radius and magnitude.
\label{fig616}}
\end{figure}

Finally, an initial estimate of the ``true'' LF and the error distribution,
magnitude biases and the total recovery probability ($F$) were used to compute
the completeness fraction $f$ (the ratio of the predicted number of stars to
the actual number of observed stars). The completeness fraction results are
shown in Figure \ref{fig616}. We then interpolated to compute $f$ for the
radial distance and magnitude of each detected star.  For each observed star
$f^{-1}$ was added to the appropriate magnitude bin to determined the observed
LF. (Note that the completeness fraction was set to 1.0 for star brighter than
the turnoff.) The observed LF along with the upper and lower 1 $\sigma$ error
bars on log $N$ are listed in Table \ref{tab51}.

\section{Discussion}

\subsection{Reddening, Metallicity, Distance Modulus, and Age}

Because NGC 5466 resides at high galactic latitude, it suffers little if any
reddening. Though \citet{sch98} found a reddening of $E(B-V) = 0.02$ from the
maps of dust IR emission, we adopted $E(B-V) = 0.0$ \citep{ros00}. For our
interests in this paper, the small difference is of small importance.
Most of the comparisons below between observations and theory are {\it
  relative}, in which reference points (like the turnoff) are used
to determine magnitude and/or color shifts. This has the benefit of minimizing
the influence of uncertainties in reddening and distance modulus (see below).

As for abundances, there is only one high-resolution measurement for
a cluster star, and it is for the anomalous Cepheid V19. \citet{mcc} find
[Fe/H]$=-1.92\pm0.05$, while \citet{pritz} find [Fe/H]$=-2.05$ using the same
data. Typically quoted metallicity values include [Fe/H] $= -2.17$
\citep{zin80} from photoelectric photometry of integrated light in selected
filter bands, and [Fe/H] $= -2.22$, which was derived by \citet{zin84}
(ultimately from low-resolution spectral scans by \citealt{sz}).  When
converted to the widely-used metallicity scale of \citet{cg97}, this becomes
[Fe/H] $= -2.14$. More work could certainly be done on the composition of NGC
5466 stars, but the evidence so far points to an abundance [Fe/H] $\la -2.0$.
Though the range in the above quoted metallicity values is relatively large
for a globular cluster, the exact value is not critical for our purposes since
we will primarily be concerned with {\it relative} comparisons.

Our photometry does not extend faint enough to derive a new distance modulus
from subdwarf fitting to the main sequence. \citet{har96} obtained
$(m-M)_{V}=16.0$ by calibrating the observed luminosity level of the
horizontal branch with the relation $M_{V}(HB) = 0.15\mbox{[Fe/H]}+0.80$ and
adopting a reddening, $E(B-V)$=0.0 and a metallicity, [Fe/H]=$-2.22$.
\citet{ferr99a} determined distance moduli $(m-M)_{V} = 16.16$ from their
zero-age HB estimate, assuming no reddening and metallicity on the
\citet{cg97} scale.  We will consider distance moduli in this range.

Most previous age estimates of NGC 5466's age have it older than the recent
determination of the age of the universe ($13.7^{+0.13}_{-0.17}$ Gyr) obtained
by the Wilkinson Microwave Anisotropy Probe (WMAP) team \citep{spe06}. Recent
homogeneous studies of GGC indicate that NGC 5466 is coeval with clusters of
similar metallicity \citep{sal02b,ros00}. As a result, we will primarily
consider ages in the range of 12 to 13 Gyr.

\subsection{The Color-Magnitude Diagram}

\begin{figure}
\plotone{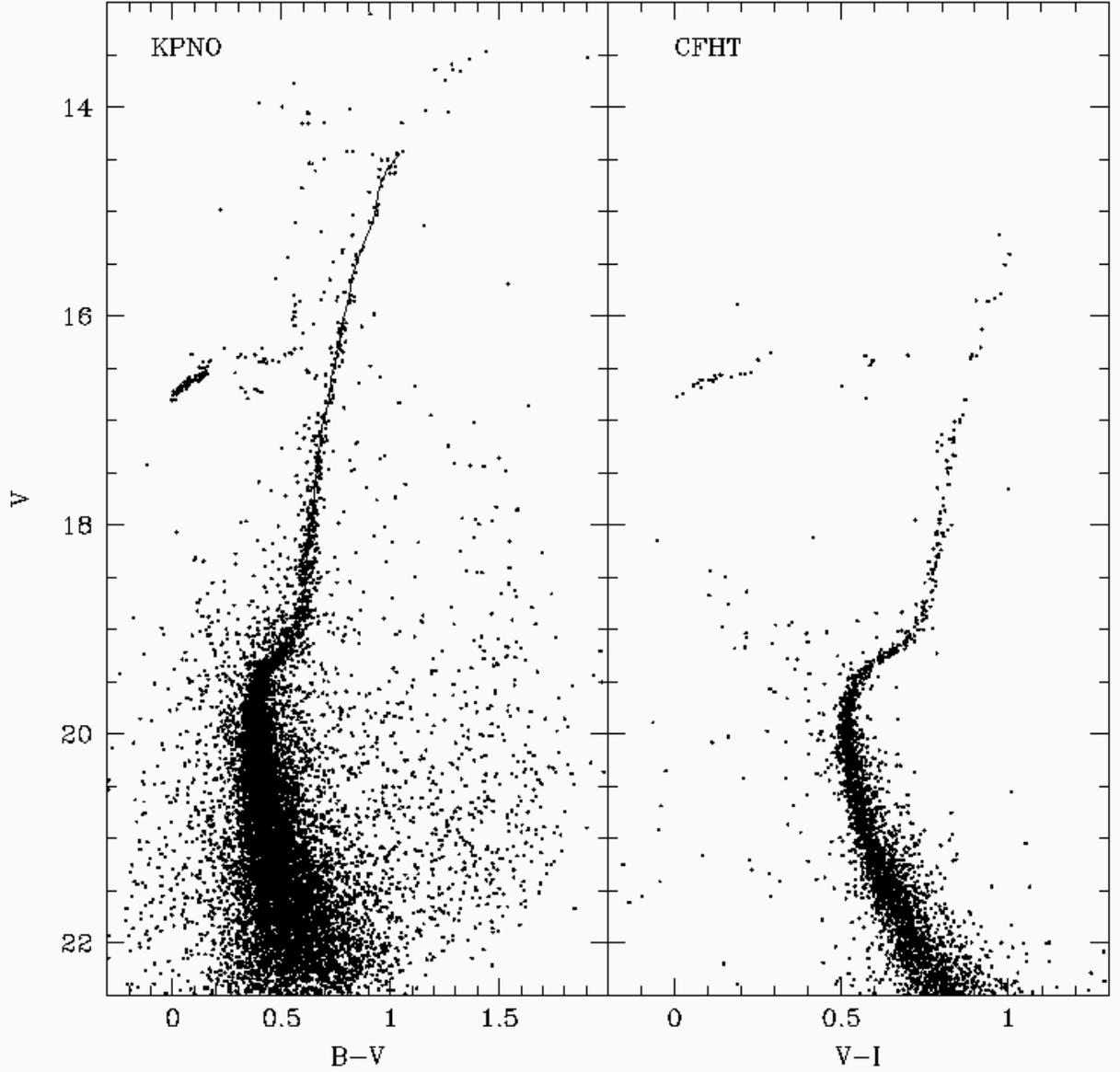}
\caption{Color-magnitude diagrams for all stars measured in the KPNO and 
  CFHT images. The $BV$ fiducial (Table \ref{tab34}) is also plotted in the
  left panel.\label{fig411}}
\end{figure}

The color-magnitude diagrams (CMDs) for NGC 5466 show well-defined RGB, AGB,
and HB sequences (see Fig. \ref{fig411}), and stars extending from the tip of
the RGB down to $V \approx 22.5$. Fiducial sequences for the MS and lower RGB
were determined from the mode of the color distribution of stars in magnitude
bins.  The SGB position was determined using the magnitude distribution of the
stars in color bins. The fiducial line for the rest of the RGB was obtained
from the mean color of stars in magnitude bins. The fiducial points are listed
in Table \ref{tab34}.

\begin{figure}
\plotone{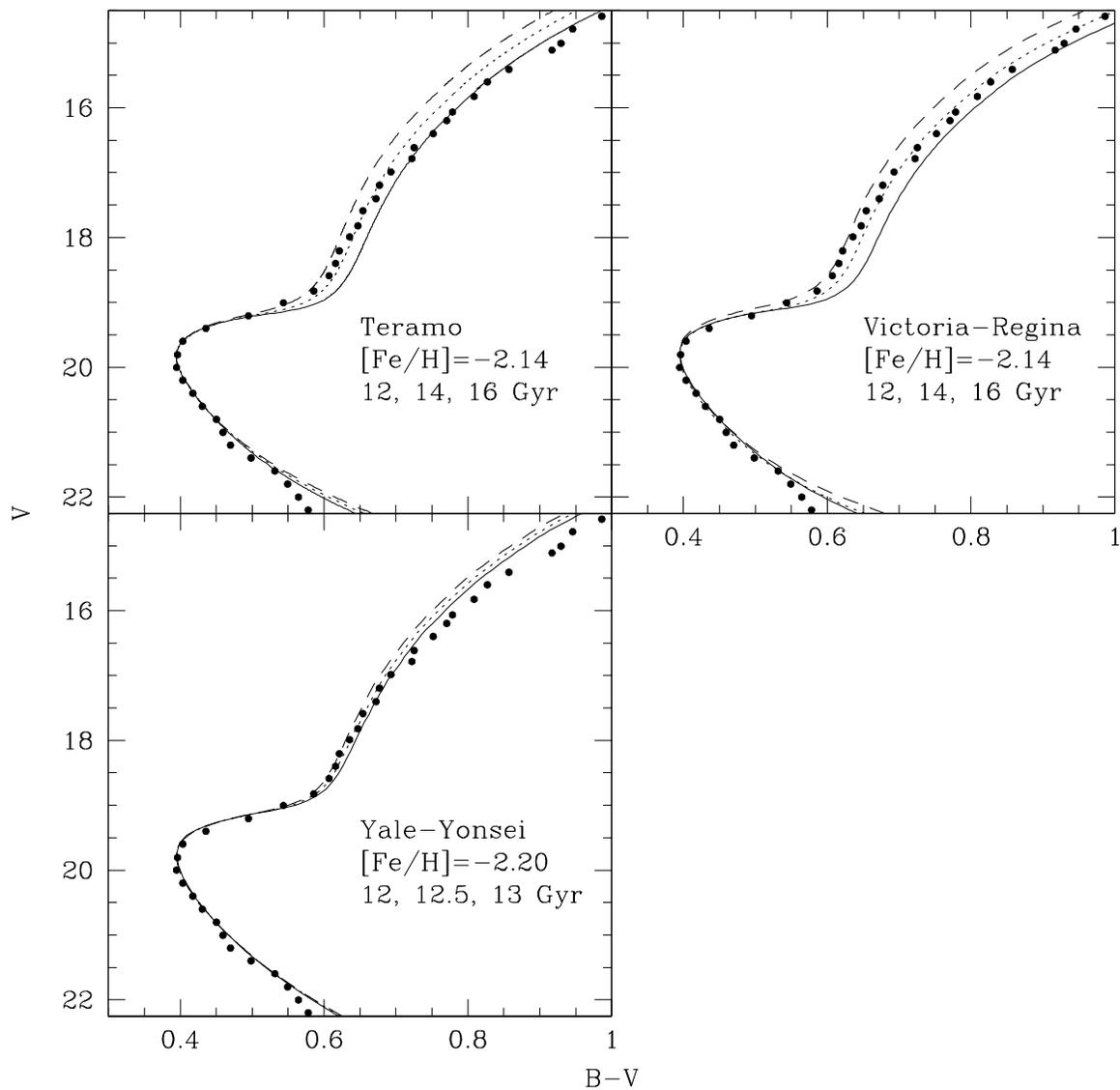}
\caption{Comparison of the observed fiducial sequence of NGC 5466 with the
isochrones of the Teramo, Victoria-Regina, and Yonsei-Yale groups. 
The isochrones have been shifted horizontally so that the turnoff colors
align, and shifted vertically to align the main sequence point 0.05 mag redder
than the turnoff. On the giant branch, the ages increase from the reddest to
the bluest isochrone.
\label{fig412}}
\end{figure}

\begin{figure}
\plotone{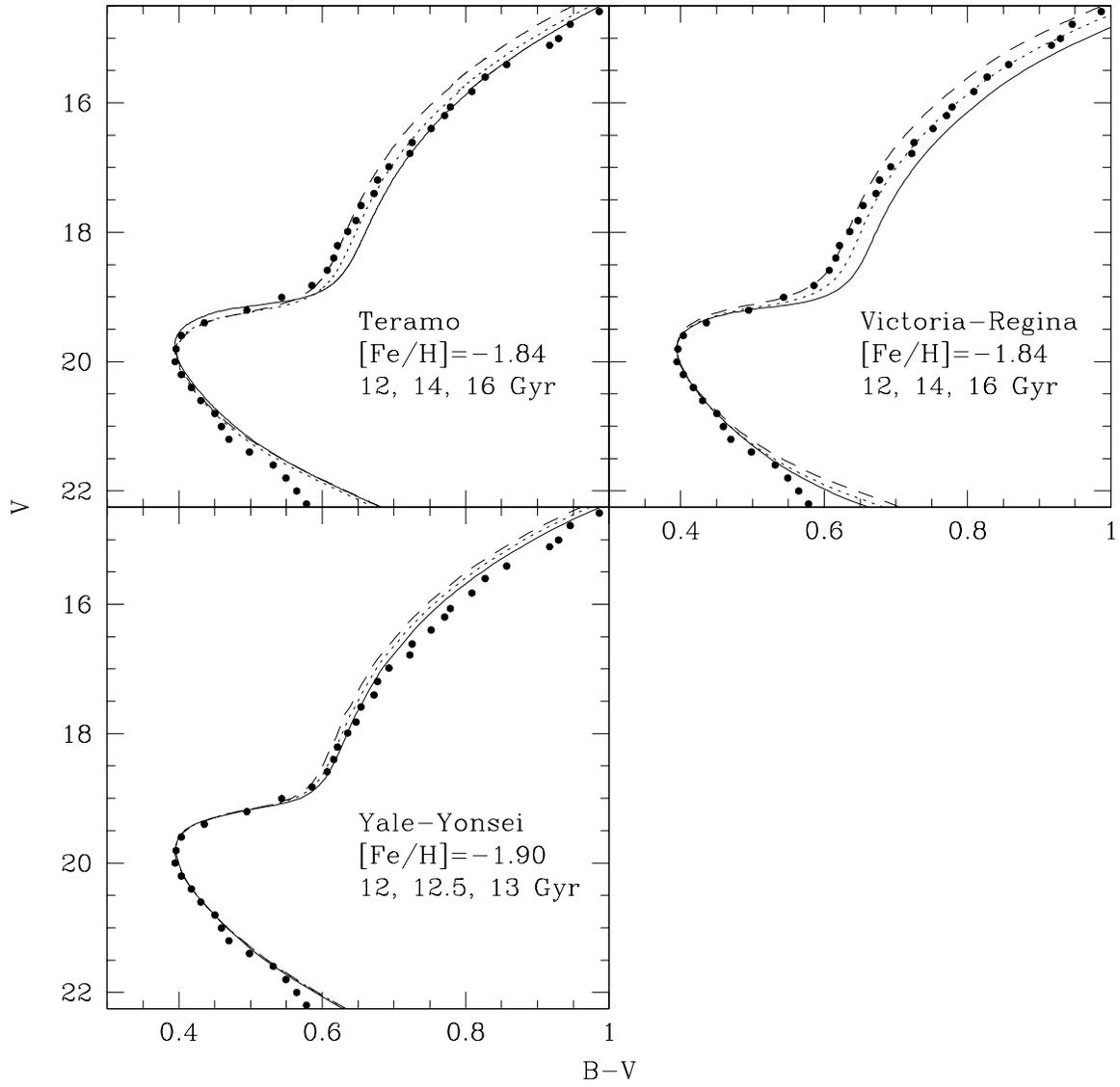}
\caption{Same as Fig. \ref{fig412}, except for more metal-rich models.
\label{rich}}
\end{figure}

A comparison of the fiducial points derived for NGC 5466 with theoretical
isochrones for a range of ages from the Teramo \citep{ter}, Victoria-Regina
\citep{vr}, and Yonsei-Yale \citep{yy} groups is displayed in Figures
\ref{fig412} and \ref{rich}.  The isochrones have been shifted in color and
magnitude (aligning the turnoff colors and the magnitudes of the main sequence
point 0.05 mag redder than the turnoff) according to the technique of
\citet{van90}. This has the advantage of removing some of systematic
uncertainties associated with the color-$T_{eff}$ transformations. [In our
comparisons, we found that the Yonsei-Yale models could not match the fiducial
line for any reasonable set of input parameters when the transformations of
\citet{lej} were used. Therefore, we only utilize models using the Green,
Demarque, \& King (1987) transformations. Even then, we could not find a match
with the slope of the upper giant branch for reasonable metallicities ([Fe/H]
$\la -1.9$).] On the whole, the shape of the fiducial matches the models well
on the main sequence, subgiant branch, and lower giant branch. Neither the
Teramo nor the Victoria-Regina models include element diffusion processes,
while the Yonsei-Yale isochrones only include He diffusion. However,
differences in $T_{eff}$-color transformations are likely to be the cause of
some of the differences seen.

\subsection{The Luminosity Function}

The number of stars at a given luminosity in post-main-sequence phases is
directly proportional to the lifetime spent at that luminosity.  It is well
known that the LF of the RGB probes the chemical stratification inside a star
because the hydrogen abundance being sampled by the thin hydrogen-fusion shell
affects the rate of evolution, and hence star counts on the RGB \citep{ren88}.
Setting aside the short pause at the RGB bump, RGB evolution accelerates
in a very regular way that is ultimately related to the structure of
degenerate core. The relationship between core mass and radius forces the
fusion shell to function at strictly controlled density and temperature
conditions, which leads to a relationship between core mass and luminosity.
This causes the LF to be particularly sensitive to certain physical details,
which we discuss in \S \ref{rgbms} below.

Fig. \ref{fig617} shows the observed luminosity function compared to
theoretical LFs for the labeled values of metallicity, exponent of the initial
mass function, and a range of age estimates for NGC 5466, assuming our
preferred distance modulus $(m-M)_V = 16.00$. The theoretical models were
normalized to the observed LF at $V \approx 21.3$ (sufficiently faint that
stellar evolution effects are minimized). The models agree well with the
observed LF, implying an age of approximately 12 - 13 Gyr. 

\subsubsection{Relative RGB and MS Numbers\label{rgbms}}

\citet{gall} recently discussed theoretical luminosity functions calculated by
different groups. One of the primary differences they noted was in the number
of giant stars relative to main sequence stars. In order to show these
differences in a parameter-independent way, we follow the method of
\citet{van90}.  In Fig. \ref{fig619}, the theoretical LFs were shifted so that
a point on the main sequence 0.05 mag redder than the turnoff color point
matched the corresponding point on the cluster fiducial line. The reason for
using the point ($V_{TO+0.05}$) rather than the MSTO itself is that the MS has
a significant slope and curvature at this point, making it possible to
accurately measure the point in both observational data and isochrones
\citep{van90}. The theoretical models were normalized to the two bins in the
observed LF on either side of the turnoff ($V = 19.83$ and 20.13). As can be
seen, age-related differences between the theoretical LFs nearly disappear
when this procedure is applied (see also \citealt{stet91a,van98}). However,
when different sets of models are compared, there are small differences in the
number of RGB stars relative to MS stars, with the Victoria-Regina models
predicting the smallest number of giants and the Yonsei-Yale models predicting
the largest.

To quantify the differences, we computed the ratio of the number of stars on
the lower giant branch to the number of stars near the main sequence.
\citet{pol05} introduced this ratio and showed that it is insensitive to age
and heavy-element abundance.  For the main sequence population, we used star
counts in the two bins on either side of the turnoff ($19.682 < V < 20.282$),
and for the red giant branch we used the counts in the range $16.982 < V <
18.482$. We derived model values from the same magnitude ranges {\it relative}
to the $V_{TO+0.05}$ point on the main sequence. Values are compared in Table
\ref{tabrat}. The error in the observed value is dominated by Poisson
statistical scatter. The Yonsei-Yale models are in best agreement with the
observations, the Victoria-Regina models are out of agreement by more
than 2 $\sigma$, and the Teramo models are in between.

It is worth examining the possible causes of this difference both because it
may help improve the physics inputs for the models and because red giant stars
are some of the largest contributors to the integrated light of old stellar
populations. \citet{gall} tabulated most of the main physics inputs for the
most widely-used model sets\footnote{Since the \citet{gall} review was
  published, the Teramo group found an error in the evolution scheme they used
  on the giant branch, which brings their models into better agreement with
  other groups. We use their updated models in the comparisons here}. Earlier
studies \citep{stet91a,van98} have shown that the LF-shifting method used above
eliminates nearly all sensitivity to model input parameters like 
mass function, age, convective mixing length, and composition
inputs with the exception of helium abundance, which we examine first.


Older models (e.g. Fig. 9 of Ratcliff 1987, Fig. 7 of Stetson 1991, Fig. 3 of
\citealt{van98}) seem to agree that an increase in {\it initial} helium
abundance $Y$ in non-diffusive models by 0.1 results in an increase in the
relative number of stars on the RGB (more precisely, a reduction in the
relative number of main sequence stars) by about 0.07-0.08 in $\log N$. The
Teramo models ($Y = 0.245$) predict about 12\% more giant stars relative to
main sequence stars compared to the Victoria-Regina models ($Y =
0.235$)\footnote{The Teramo models also assume a larger $\alpha$-element
  enhancement ([$\alpha$/Fe] = 0.4) than the Victoria-Regina models
  ([$\alpha$/Fe] = 0.3), which would tend to {\it reduce} the number of giants
  relative to main sequence stars. However, because the relative number of RGB
  and MS stars is not sensitive to small differences in heavy element
  abundance, this difference is probably unimportant.}. This difference
corresponds to a shift of 0.05 in $\log N$, which is about an order of
magnitude too large for the helium abundance difference.

The Yonsei-Yale models have the lowest assumed helium abundance ($Y = 0.23$),
but are the only set of the three that include helium diffusion.  The
inclusion of helium diffusion reduces the age derived from the turnoff of a
globular cluster by about 10-15\% \citep{proff,scl,van02}, thanks to the
inward motion of helium.  According to theoretical models (e.g.  Fig.  8 of
\citealt{proff}), diffusion has a small effect on the LFs ($\sim$ a few times
$10^{-2}$ in $\log N$, increasing for increasing age), but it does increase
the number of giants relative to MS stars.  He diffusion reduces the total
core hydrogen fuel supply available to an MS star, but in itself this does not
strongly modify the LF, just changes the brightness of the turnoff. This
magnitude change is eliminated in our LF shifting procedure. The chemical
composition profile left in the star after it leaves the main sequence has a
greater impact. Diffusion reduces the H abundance in the fusion regions,
thereby decreasing the evolutionary timescale. According to the models of
\citet{pm}, the changes to the He abundance profile are most considerable
immediately below the surface convection zone, and just outside the nuclear
fusion regions (where the composition gradient slows the inward settling of
helium). However, over most of the star, the changes in $Y$ are limited to
0.01 - 0.02. Because the core portion of the composition profile is consumed
on the subgiant branch, the evolution timescales for giant stars are only
affected in a minor way, and the appearance of a deep convection zone on the
lower giant branch wipes out most of the effects of diffusion for the upper
giant branch evolution.

In spite of this, the Yonsei-Yale models have almost 23\% more giants than the
Victoria-Regina models, and more than 9\% more giants than the Teramo models
(relative to MS stars). The lower helium abundance in the Yonsei-Yale models
compared to the other models should partially counteract what effects helium
diffusion might have had on the RGB/MS ratio as well. Thus, it appears that
neither He abundance nor He diffusion can completely explain the differences
between the Yonsei-Yale models and the other groups.

We can ask whether the LFs show similar disagreements at other metallicities.
\citet{harg04} made comparisons between the Victoria-Regina and Yonsei-Yale
theoretical LFs and observational LFs for the clusters M3 \citep{rood99}, M5
\citep{san96}, M12, and M30.  The overall impression from those comparisons
was again that the Yonsei-Yale models (having He diffusion) predict more giant
stars relative to main-sequence stars than do the Victoria-Regina models.  In
Fig. \ref{otherlfs}, we compare the LFs for these clusters with the Teramo
models. The degree of agreement or disagreement can be quantified with number
ratios of lower RGB and MSTO stars, similar to the ones we computed earlier
for comparisons with NGC 5466.  Our calculations are shown in Table
\ref{tabrat}. As can be seen, uncertainties in the metallicity scale have some
effect on the comparisons with the observations. The \citet{cg97} scale has
higher [Fe/H] values than the \citet{zin84} scale, and thus results in lower
number ratios.

Fig. \ref{rats} shows the results of comparing the observed ratios
with the models for different [Fe/H] scales.  On the \citealt{zin84} scale
(right panels), the observed values seem to be in agreement to within about $1
- 1.5 \sigma$ for most of the models, with the exception of the the lowest
metallicity clusters (M30 and NGC 5466) and the lowest helium abundance
(Victoria-Regina) models. On the \citealt{cg97} scale, the Yonsei-Yale models
have the best overall agreement, although the Teramo models only deviate
noticeably for the lowest metallicity clusters. The Victoria-Regina models
predict too few giants for all of the clusters.

The differences from model to model (as opposed to models versus observation)
point toward deficiencies elsewhere in the physics or computational algorithms
used in the stellar evolution codes. \footnote{For a comparison of these
  physical inputs for the different theoretical groups, see Table 1 of
  \citet{gall}.}  The RGB LF is a robust prediction of the models because
there is a strong core mass --- luminosity relationship: the conditions in the
hydrogen fusion shell of the giant are strongly dependent on the structure of
the degenerate core and are almost independent of the details of the mass or
structure of the envelope. As a result of this, we can focus on factors
affecting core structure.  (As a non-standard physics example, \citealt{van98}
describe the way in which core rotation relieves a giant star of some of the
need to support itself by gas pressure, which reduces the core temperature and
lengthens the evolutionary timescale.) Because model-to-model differences
appear even on the faint end of the giant branch, we can set aside factors
that only become important to the structure of the star near the tip of the
giant branch [such as neutrino losses and conductive opacities; see
\citealt{bjork}, for example], even though there are significant theoretical
uncertainties in these quantities. Nuclear reaction rates in the fusion shell
can also be neglected, partly because the uncertainties in the reactions
appear to be relatively small \citet{adel}, but also because small changes in
the reaction rates require only tiny changes in the shell temperature to get
the same energy production. This leads us to examine the equation of state
(EoS) in the core.

Although the behavior of degenerate electrons is thought to be very well
understood, their interactions with nuclei can have a measureable effect on
the pressure. Particles of like charge tend to cluster together, which
modifies the free energy of the gas and reduces the gas pressure for given
density and temperature. \citet{hk} looked at the effects of the inclusion of
Coulomb interactions on giant stars, and their results are corroborated by
those of \citet{cass}. They found that for a given core mass the fusion shell
temperature was higher when the Coulomb interactions were included, which
leads to faster processing of hydrogen.  Thus this is another example (like
core rotation) where modification of the pressure support of the core affects
the evolutionary timescales, which results in changes to the luminosity
function. The Coulomb corrections to the pressure become more important with
increasing density for the core, but are small compared to the contribution of
the degenerate electrons.

All of the model sets we have considered here incorporate Coulomb interactions
in some form.  The Teramo group used the most sophisticated ``EOS1'' version
of the FreeEOS\footnote{FreeEOS is available at
  http://freeeos.sourceforge.net/, and the discussion of the implementation of
  the Coulomb effect can be found at
  http://freeeos.sourceforge.net/coulomb.pdf.}, which incorporates Coulomb
corrections in a form that matches limits in both the weak (Debye-H\"{u}ckel)
and strong (once-component plasma) Coulomb interaction limits as well as (less
importantly) electron exchange interactions. The strong interaction limit is
most relevant for giant star cores since the strong interaction parameter
\[ \Gamma = \frac{\zeta^2 e^2}{r_0 kT} \ga 1, \]
where $\zeta$ is the rms nuclear charge and $r_0$ is the average internuclear
distance. 


The Yonsei-Yale group used the OPAL EoS tables \citep{opaleos}, but falls back
on the group's older EoS [e.g.  \citep{gue}] for conditions for high densities
and temperatures outside the OPAL tables. While the most recent OPAL tables
probably contain the most complete physical description of the Coulomb effect,
the OPAL tables they used (Y.-C. Kim, private communication) were computed
prior to recent improvements to account for relativistic electrons \citep{rn},
and as a result cut off at $\log \rho > 5.0$. The Yale EoS at higher densities
{\it only} includes the Coulomb effect in the weak Debye-H\"{u}ckel limit,
which  for the highest densities in the core. The Victoria-Regina models
also use a modified version of the EFF EoS \citep{eff}, with a correction for
Coulomb corrections in the weak Debye-H\"{u}ckel limit \citep{vsria}.

The differences in the implementation of the Coulomb effect may explain the
fact that the Teramo models generally predict more giants (relative to the
main sequence) than the Victoria-Regina models do. However, the smaller Coulomb
corrections in the Yonsei-Yale models would tend to result in fewer giants
than the Teramo models (although the effects of helium diffusion work in the
opposite direction). So, we are unable to completely reconcile the differences
in the luminosity functions from the three groups. 

Obviously more detailed study is needed by all of the modelling groups to
identify the causes of the differences, but such a study is beyond the scope
of this paper. Still, we believe that helium diffusion and strong interaction
Coulomb corrections are physical effects that should be considered first.
There is, for example, good evidence from
helioseismology for helium diffusion in the Sun \citet{bpw}, despite the
surface convection and meridional flow (e.g., \citealt{hath}). It is expected
that helium diffusion should also act in globular cluster stars.
A detailed study of the effect of equation of state uncertainties has yet to
be done [see, for example, \citet{bjork} for a study of uncertainties in other
physical inputs]. Use of FreeEoS would make a study of equation of state
effects most stringent since it appears to be capable of modelling the most
sophisticated tabular EoS (OPAL), while also having the flexibility to allow
individual bits of the physics to be ``turned off''.



As a final warning about the observations, we should remember the LF of the
cluster M10.  \citet{pol05} found that unusual variations in numbers of RGB
stars at different brightness levels in M10 (a virtual twin to M12). In
particular there seemed to a significant excess in the number of stars near
the RGB bump in brightness, while the lower RGB appeared normal (compared to
Victoria-Regina and Yonsei-Yale models). A similar excess may be present in
the RGB LF of M13 \citep{cho}. These kinds of variations cannot be explained
by the ``global'' physics that should apply to all globular cluster stars.
These anomalies point toward fluctuations in the stellar initial mass function
or composition-dependent effects.

\subsubsection{The RGB Bump\label{bumpsec}}

A second feature of the LF presented here is a noticeable RGB bump.  Typically,
the RGB bump appears as a peak in the differential LF and as a change of slope
in the cumulative LF (CLF). The bump provides a measure of the maximum depth
reached by the outer convection zone during first dredge-up since it is the
result of a pause in the star's evolution when the shell fusion source begins
consuming material of constant, lower helium content \citep{fus90}.
Unfortunately, the number of stars occupying the bump gets smaller and the
luminosity of the bump increases as the metallicity of the cluster decreases,
making the bump harder to detect in metal-poor clusters. A small peak appears
in our differential LF at $V \approx 16.2$, and a
significant ($2.5 - 3.5 \sigma$) change in slope occurs at the same position
as the peak in the differential LF, as shown in Fig. \ref{fig621}.

The relative brightness of the bump can be measured by comparing to
the $V$-magnitude of the HB at the level of the RR Lyrae
instability strip $\Delta V^{bump}_{HB} = V_{bump}-V_{HB}$ \citep{fer99}.
This indicator is a function of the total metallicity and the age of the
cluster: an increase in metallicity and/or a decrease in age are accompanied
by a decrease in luminosity of the bump \citep{fer00}.
We find
$V_{bump} = 16.20\pm0.05$ mag, 
and $V_{HB}=16.52\pm0.11$ (from interpolation
between the average magnitudes of non-variable HB stars at the
blue and red ends of the RR Lyrae instability strip), giving $\Delta
V^{bump}_{HB} = -0.32\pm0.12$. 
In the compilation of \citet{fer99}, a zero-age HB reference point was
calculated using the relation $V_{ZAHB} = V_{HB}
+0.106\mbox{[Fe/H]}^2+0.236\mbox{[Fe/H]}+0.193$. \citeauthor{fer99} found
$V_{ZAHB}=16.62\pm0.10$, which is consistent with the value obtained here
($V_{ZAHB}=16.65\pm0.11$). Our value of $\Delta V_{ZAHB}^{bump} = -0.45 \pm
0.12$ is considerably lower than tabulated values for other clusters with
similar metallicities (M68: $-0.60\pm0.07$; M92: $-0.65\pm0.12$; M15:
$-0.65\pm0.09$). In a separate compilation, \citet{zoc99} measured a smaller
value $\Delta V_{ZAHB}^{bump} = -0.45 \pm 0.11$ for M15, in better agreement
with the value for NGC 5466. (The difference is primarily because
\citeauthor{zoc99} measured the bump position to be 0.16 mag fainter than
\citeauthor{fer99}.) Clearly there is still some need for more precise
comparisons of bumps in metal-poors clusters with theory.


We believe, however, that the brightness of the bump should ultimately be
judged using {\it hydrogen-fusing} stars as references because it avoids any
effects of the poorly-understood physical processes (such as the helium flash
and/or mass loss) associated with the creation of an HB star. Using the
cluster LF (as seen in Figure \ref{fig619}), we again find that the observed
RGB bump is fainter than model values by at least 0.3 mag when the models are
shifted to match the cluster's main sequence. \citet{harg04} did similar
comparisons between theoretical models and luminosity functions for M3, M5,
M12, and M30 taken from the literature.  With the exception of M30 (because
the bump could not be identified), the position of the bump relative to the
turnoff region agreed well with theory.  NGC 5466 is thus the most metal-poor
cluster this comparison has been done for. So at present we are left with the
question of whether this might result from the cluster's low metallicity, or
whether we have been the unfortunate victims of a fluctuation in the number of
giant stars in this low-mass cluster. We therefore encourage the
examination of the luminosity function of more massive metal-poor clusters to
settle the question.

\subsubsection{Mass Function Exponent}


Two recent papers \citep{belo,grill} reported the discovery of tidal streams
covering many degrees around NGC 5466 in Sloan Digital Sky Survey images.
\citet{gne} examined the Milky Way globular clusters and found NGC 5466 to be
a cluster that has probably been strongly affected by disk shocking in the
recent past.  In our examination of the LF, we found a rather low value for
the global main sequence mass function slope.  The mass function for a cluster
is typically expressed as a power law ($N(M) \propto M^{-(1+x)}$), where the
slope $x=1.35$ is the standard Salpeter value.  Generally, the present-day
power-law index $x$ varies from cluster to cluster.  The mass function slopes
that best fit the upper LF of NGC 5466 around and above the MSTO have $-1 \la
x \la 0$. (Note that the best fit slope does depend on the models being used:
the Yonsei-Yale models require a flatter slope than the Victoria-Regina and
Teramo models.) Such a shallow mass function slope is unusual for a metal-poor
cluster. For example, the cluster NGC 5053 has similar metallicity, position
relative to the galactic center and plane, and density structure, but still
has a steep $x \sim 2$ mass function \citep{fahl}.  \citet{djo93} found that
mass function slopes in the range of $0.5\leq\frac{M}{M_{\sun}}\leq0.8$ are
influenced primarily by the cluster's position in the galaxy, and to some
extent by cluster metallicity. Based on both of those factors, NGC 5466 should
have a larger mass function slope ($x \sim 3$ according to the multivariate
formula in \citealt{djo93}). Like other halo clusters, NGC 5466's orbit is
quite eccentric and will take the cluster more than 30 kpc away from the
Galactic center \citep{dines}, but it is currently on its way back into the
halo after two relatively recent passes through the Galactic disk. Recent
losses of low-mass stars may explain the recent identification of strong tidal
tails near the cluster by \citet{belo}.

\section{Blue Stragglers}\label{bss}

Blue stragglers (BSSs) were first identified by \citet{san53} in the globular
cluster M3. These stars are more massive than the turnoff mass and occupy the
space in the CMD just bluer and brighter than the MSTO. Blue stragglers are
found in clusters, and relatively more frequently in lower-luminosity clusters
\citep{fer93,ps00,pio04,san05}.  From the various models proposed for the
formation mechanism of BSSs, the ``collision'' theory (involving strong
gravitational interactions between previously unassociated single or binary
stars) and the ``mass-transfer'' theory (in which the more massive star in a
binary evolves and during its expansion transfers mass to its companion) are
the strongest possibilities. There is a continuing interest in the study of
BSSs because they may provide insight into the recent dynamical history of a
cluster.

In order to identify BSSs over the entire observed area of the cluster
out to a radius of 11\farcm6, we used photometry from three datasets. In the
core of the cluster we used the CFHT data presented here for the first time.
Outside of the CFHT field, we used the $BV$ photometry of \citealt{jeo04},
which covered a field 11\farcm6 on a side centered on the cluster. Finally, we
used our KPNO data for the least crowded outskirts of the cluster. Even though
NGC 5466 is a very low density cluster, the spatial resolution of the KPNO
data was such that blends of stars would have resulted in the spurious
identification of 10 objects as BSS in the intermediate portion of the
field.

48 BSS candidates were identified in NGC 5466 by \citet{nem87}, 
all located between
$0\farcm1$ and $5\farcm6$ from the cluster center. 
In spite of the low cluster density, we find new
BSS candidates at all radii and luminosity levels, and find several of
their candidates are spurious.  According to the CFHT photometry, the object
with ID 45 from \citeauthor{nem87} is a blend of several fainter
stars, none of which is a BSS. In addition, IDs 6 and 24 were identified
as blends of stars using the \citet{jeo04} dataset. Our BSS list is
presented in Table \ref{tab41}. The list includes the nine known SX Phoenicis
stars (ID 27, 29, 35, 38, 39, and 49, \citealt{nm90}; ID 3 (SX Phe 3), 36 (SX
Phe 2), and 50 (SX Phe 1), \citealt{jeo04}) and the three eclipsing binaries
(ID 19, 30, and 31; \citealt{mat90}). New straggler candidates were given ID
numbers that build upon the \citet{nem87} list. Figure \ref{fig555} shows the
CMDs used to select the 94 identified BSSs in each of the three datasets.

In order to use the BSSs to constrain cluster dynamics, we compared the
normalized cumulative radial distribution of the BSSs to the
population of the giant branch, as shown in Fig. \ref{fig511}. \citet{nem87}
found a 97.8\% probability that their BSS sample was more centrally
concentrated than red giants in the same magnitude range. Because their
photometry was taken in conditions of poorer seeing (compared to our CFHT
photometry and that of \citeauthor{jeo04}), their samples are likely to be
somewhat incomplete near the cluster center.

Our RGB sample contains 350 stars with magnitude $V < 18.5$.
Kolmogorov-Smirnov (K$-$S) probability tests were used to test the hypothesis
that both populations were drawn from the same parent population. The K$-$S
probability that the BSSs are drawn from the same radial distribution as the
RGB population is $8.1 \times 10^{-7}$, and $2.4 \times 10^{-4}$ for the
comparison with the HB population.  By contrast there is a probability of 0.27
that the RGB and HB samples are drawn from the same population. The
concentration of BSSs toward the cluster center as compared to the RGB and HB
samples is consistent with the idea that they are more massive than individual
RGB stars, and as a result have been segregated by mass deeper within the
cluster potential well.

\citet{pio04} recently used samples of stragglers from the cores of 56
globular clusters to show that there was a strong correlation between
$F^{HB}_{BSS} = N_{BSS} / N_{HB}$ and integrated cluster $V$ magnitude, and a
weaker anti-correlation with central density. \citet{san05} examined an
additional 13 low-luminosity globular clusters using similar selection
criteria. NGC 5466 is an interesting cluster in relation to these samples
because it has an integrated luminosity that puts it at the faint end of the
Piotto et al. sample ($M_{V_t} = -6.96$; \citealt{har96}), but with a central
density that is nearly an order of magnitude lower than any of their clusters
[$\log (\rho_0 / (L_{V,\odot} \mbox{pc}^{-3})) = 0.88$] but comparable to
clusters in the Sandquist sample. To put NGC 5466 in the context of the Piotto
et al. and Sandquist samples, we selected a subset of our BSSs that
satisfied the selection criteria in those studies (brighter than the MSTO, and
bluer than the MSTO by 0.05 in $B-V$ color). From mode fitting to the turnoff
region in the \citet{jeo04} and CFHT data, we find $(B-V)_{TO} = 0.367$ and
$(V-I)_{TO} = 0.511$. A color offset of 0.05 in $B-V$ corresponds to an offset
of about 0.075 in $V-I$ \citep{vandc} for NGC 5466's metallicity. We find that
75 BSSs are brighter than the cluster turnoff ($V_{TO} = 19.99 \pm
0.05$) and 0.05 bluer than the cluster turnoff in $B-V$.  We have identified
97 HB stars in our datasets for NGC 5466, which gives a specific frequency
$F^{HB}_{BSS} = 0.77 \pm 0.12$ (with the error estimate from Poisson
statistics).


When compared to the \citeauthor{pio04} values (see Fig. \ref{fbss}), NGC 5466
falls within the general trend versus $M_{V_t}$ despite the cluster's low
central density. On the other hand, NGC 5466 has a lower $F^{HB}_{BSS}$ value
than other clusters of similar central density (but lower total luminosity).
As discussed by \citet{san05}, this provides additional evidence that the
plateau in $F^{HB}_{BSS}$ seen for clusters with $\log \rho_0 \la 2.5$ is a
result of the correlation between cluster integrated magnitude and central
density. The lowest luminosity clusters in the Sandquist sample (E3 and
Palomar 13) have central densities comparable to that of NGC 5466, but
BSS frequencies that are several times higher. Another
moderate-luminosity, low-density cluster (NGC 5053; Hiner et al., in
preparation) similar to NGC 5466 shows a comparably low straggler frequency.
BSSs produced via purely collisional means are not likely to show this
kind of behavior. More likely is the scenario proposed by \citet{dav04} in
which binary stars that would normally produce BSSs are destroyed
earlier in the cluster's history. More direct observational support for that
hypothesis is needed though --- for example, a detailed study of the variation
of the binary star fractions as a function of integrated cluster magnitude.




Fig. \ref{fig512} shows that the number and frequency 
\[ R_{BSS} = \frac{N_{BSS} / N_{BSS}^{tot}}{L^{sample}/L^{sample}_{tot}}\]
of BSSs 
relative to the integrated $V$-band flux (derived from a King model profile)
as a function of radius. Both frequencies increase toward the cluster center,
and neither shows signs of rising toward larger radii.  As recent studies of
denser clusters show (M3: \citealt{fer93}; M55: \citealt{zag97}; 47 Tuc:
\citealt{fer04}; NGC 6752: \citealt{sab04}), the BSS frequency generally
decreases at intermediate radii and rises again at larger distances. However,
the cluster Palomar 13 \citep{cla04}, which has a central density similar to
NGC 5466, shows no sign of an increase in straggler frequency at large
distance. 

In more massive clusters, the minimum in the BSS frequency is reached
approximately where the timescale for dynamical friction equals the age of the
cluster \citep{warren}. A similar calculation for the current structure of NGC
5466 indicates that this occurs at about 270\arcsec (about $2.8 r_c$). This
appears to be in the outer reaches of the core straggler distribution. Because
it is likely that NGC 5466 has lost a significant fraction of its mass, we
expect that the current density structure of the cluster has not existed
throughout its history and that NGC 5466 might have been able to dynamically
relax stragglers to its core from larger distances earlier in its history.
This may be showing that the global BSS population differs significantly
between low-density/low-mass clusters and high-density/high-mass clusters. A
lack of stragglers at large distance may be signature of large-scale tidal
stripping of the cluster, which would remove both stragglers that would
normally have formed in primordial binaries in the outer reaches of the
cluster and ones that formed in the core but were given velocity kicks into
orbits that would take them into the outer reaches.

The case against a rise at large radius in Palomar 13 is stronger because the
cluster has been surveyed out to 19 core radii, while in NGC 5466, we have
only surveyed out to about 10 core radii (or 7.5 half-mass radii). Still, NGC
5466 probably should have an even more concentrated distribution of stragglers
if its current density structure has existed for most of its history. In more
massive clusters, the secondary rise in straggler frequency is observed
between 8 and 10 $r_c$.  Unfortunately, further study of the stragglers in NGC
5466 will probably be complicated by the strong tidal tails observed in the
cluster.


\section{Conclusions}

Examinations of the luminosity functions of globular clusters continue to
produce interesting tests of astrophysics. In this study, we found that NGC
5466 has a luminosity function that is in better overall agreement with
theoretical models than the anomalous cluster M30, which has a similar low
metallicity.  In addition, we found that the relative numbers of red giant and
main sequence stars may produce a fairly sensitive test of the physics near
the core of red giants --- specifically, helium diffusion and Coulomb
interactions. However, we are not yet able to fully explain the differences
between sets of theoretical models.

Recent discoveries of large tidal tails associated with NGC 5466 suggest that
this cluster has been strongly disrupted by interactions with the Galaxy. Our
measured flat ($-1 \la x \la 0$) main-sequence luminosity function is unusual
for a low-metallicity halo cluster. It is, however, consistent with the
emerging picture of mass-segregation followed by tidal stripping.

We have thoroughly re-examined the blue straggler population in the cluster,
and detected a total of 94. The radial distribution of stragglers is
clearly more centrally concentrated than the RGB and HB populations.
The frequency of blue stragglers in the cluster is relatively low ---
consistent with the observed anti-correlation between frequency and
cluster luminosity, in spite of the cluster's very low central
density.

\acknowledgments We would like to thank the anonymous referee for helpful
comments on the manuscript, Y. Jeon for providing us with an electronic copy
of his photometric dataset, Y.-C. Kim for information on the Yonsei-Yale
isochrones, and S. Cassisi for providing us with access to the Teramo set of
models. This work has been funded through grants AST 00-98696 and 05-07785
from the National Science Foundation to E.L.S. and M.B.

\clearpage

\clearpage

\clearpage

\clearpage

\clearpage

\clearpage

\clearpage

\clearpage

\clearpage

\clearpage

\clearpage

\clearpage

\clearpage
\begin{figure}
\plotone{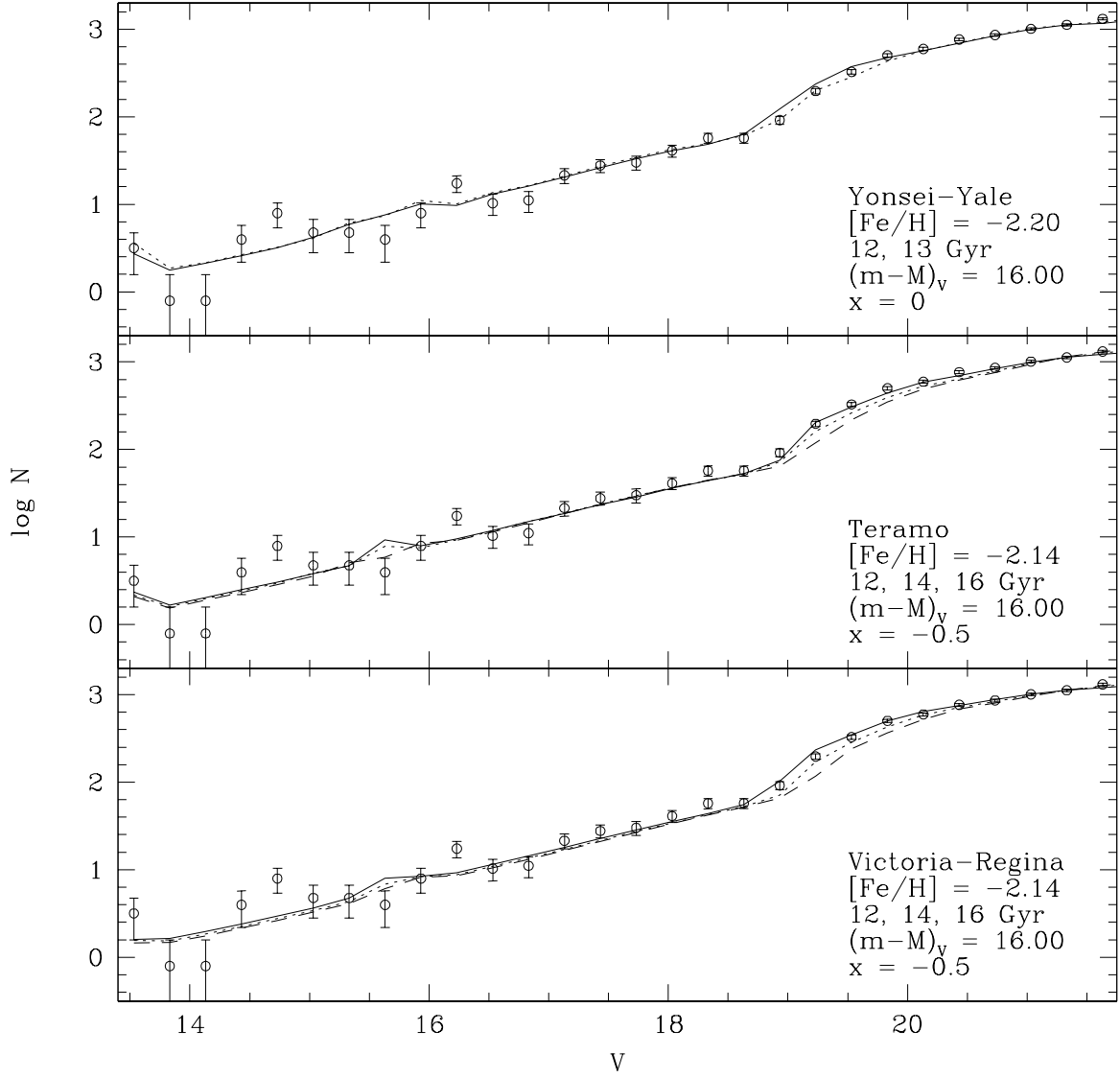}
\caption{Comparison of the observed $V-$band LF of
  NGC 5466 with theoretical models of the Yonsei-Yale,
Teramo, and Victoria-Regina groups assuming $(m-M)_V=16$.
\label{fig617}}
\end{figure}

\clearpage
\begin{figure}
\plotone{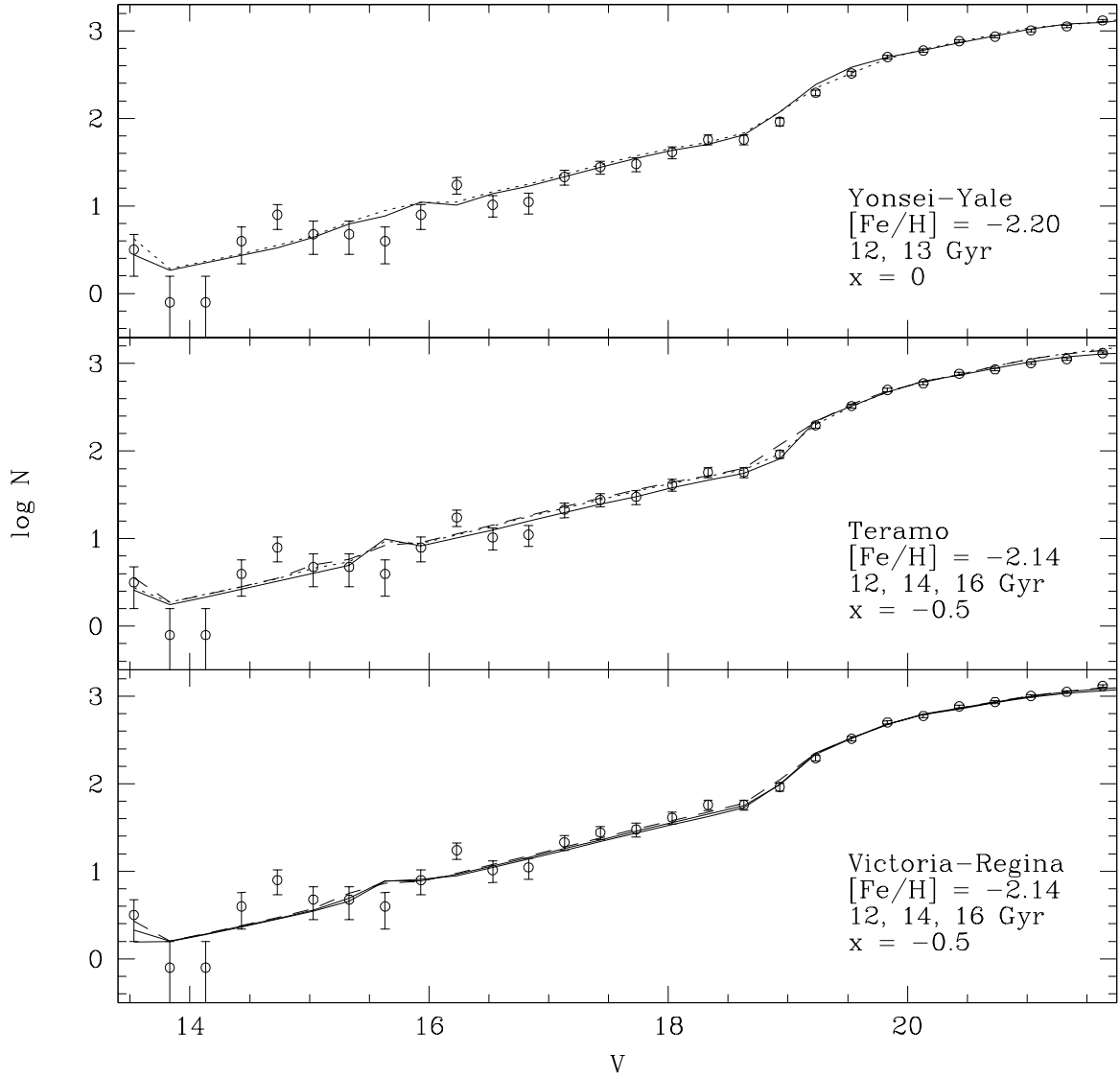}
\caption{Comparison of the observed $V-$band LF of
  NGC 5466 with theoretical models of the Victoria-Regina, Yonsei-Yale, and
  Teramo groups using magnitude shifts that bring the main sequence point 0.05
  mag redder than the turnoff into alignment. The models have been normalized
to the two bins on either side of the turnoff ($V = 19.94$).
\label{fig619}}
\end{figure}

\clearpage
\begin{figure}
\plotone{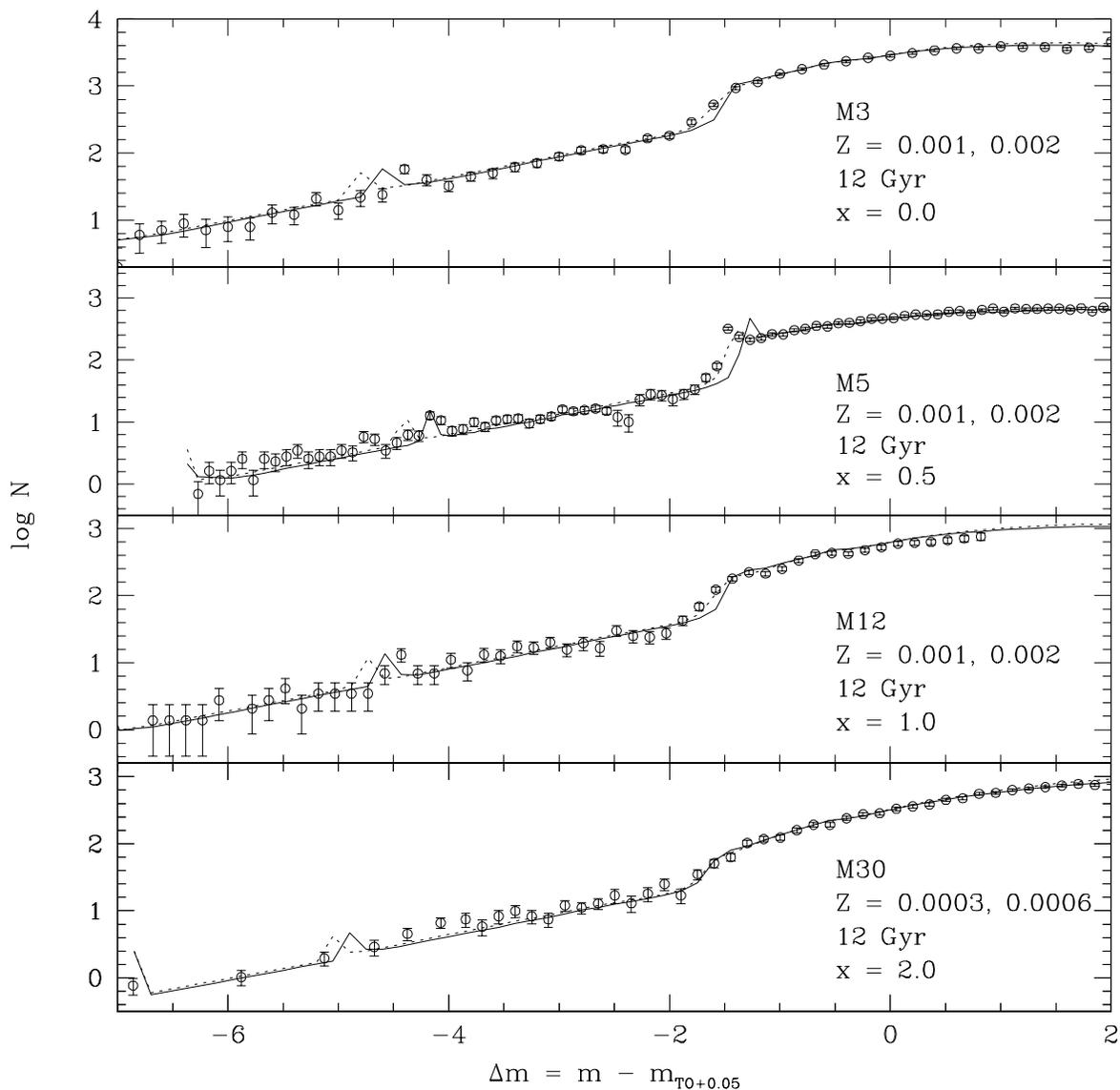}
\caption{Comparison of the observed LFs of
  M3, M5, M12, and M30 with theoretical models of the Teramo group using
  magnitude shifts that bring the main sequence point 0.05 mag redder than the
  turnoff into alignment. The models have been normalized to bins on either
  side of the turnoff.
\label{otherlfs}}
\end{figure}

\clearpage
\begin{figure}
\plotone{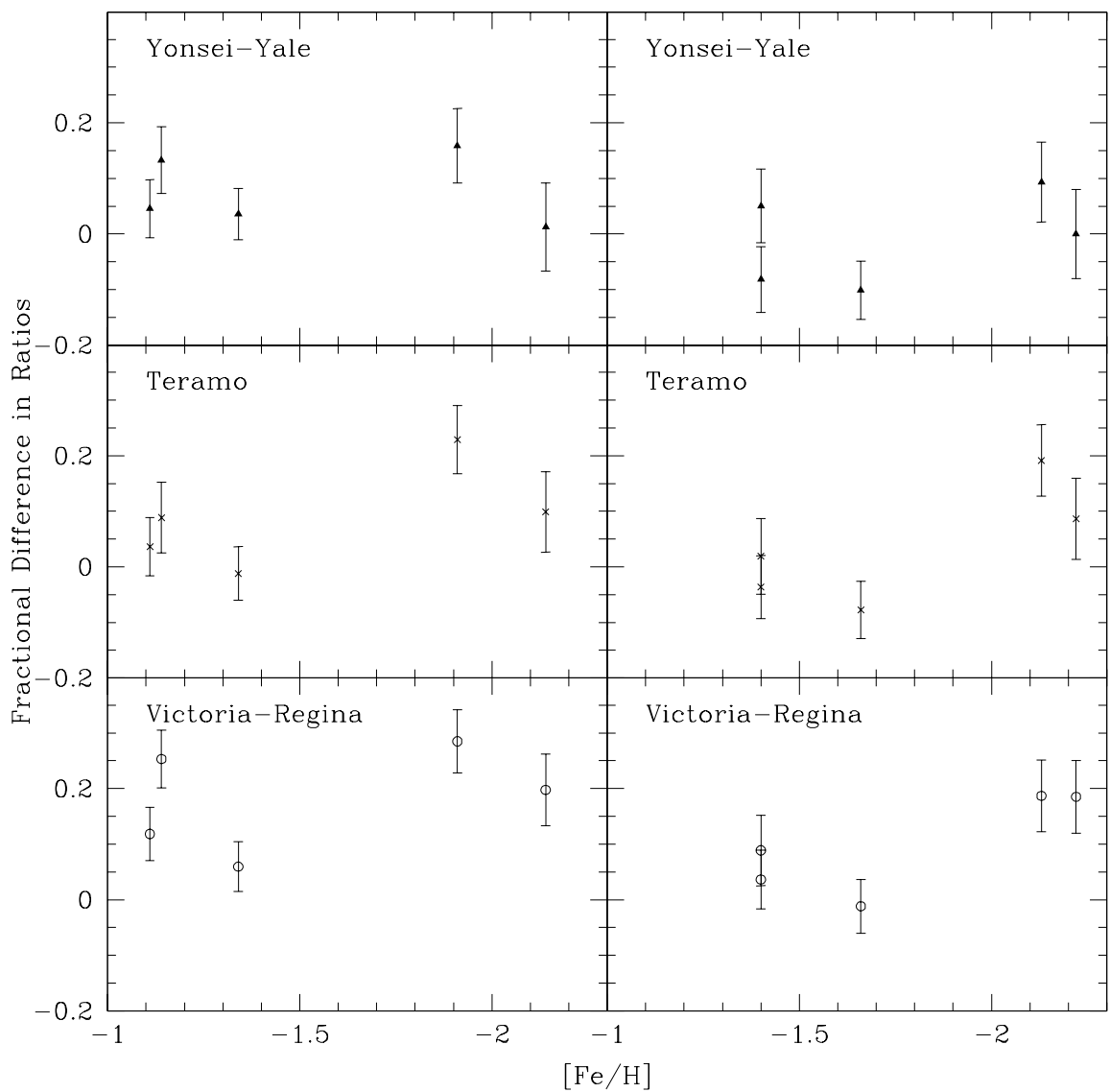}
\caption{Fractional difference between the observed RGB-MS number ratios
(for M5, M12, M3, M30, and NGC 5466, from left to right) and the theoretical
  predictions from the Yonsei-Yale, Teramo, and Victoria-Regina models. The
  left panels use the \citet{cg97} metallicity scale, and the right panels use
the \citet{zin84} scale. The sense is (observed $-$ theoretical) / observed.
\label{rats}}
\end{figure}

\clearpage
\begin{figure}
\plotone{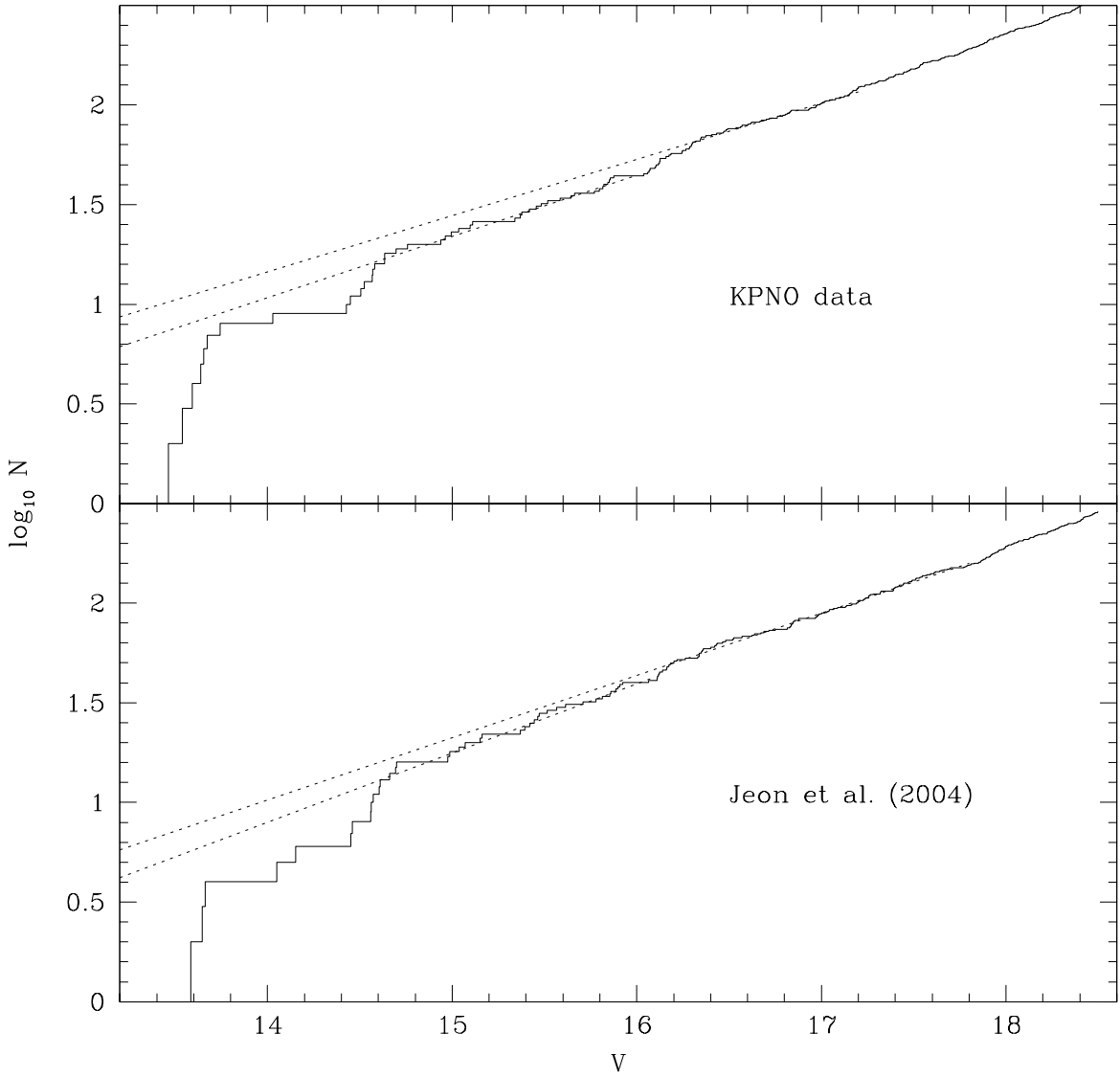}
\caption{The cumulative luminosity function for bright RGB stars derived from
  the photometry of Jeon et al. (2004; 286 stars) and from the KPNO dataset
  (338 stars) presented here. Dotted lines show fits to the data for stars
  above and below the position of the apparent bump.
\label{fig621}}
\end{figure}

\clearpage
\begin{figure}
\plotone{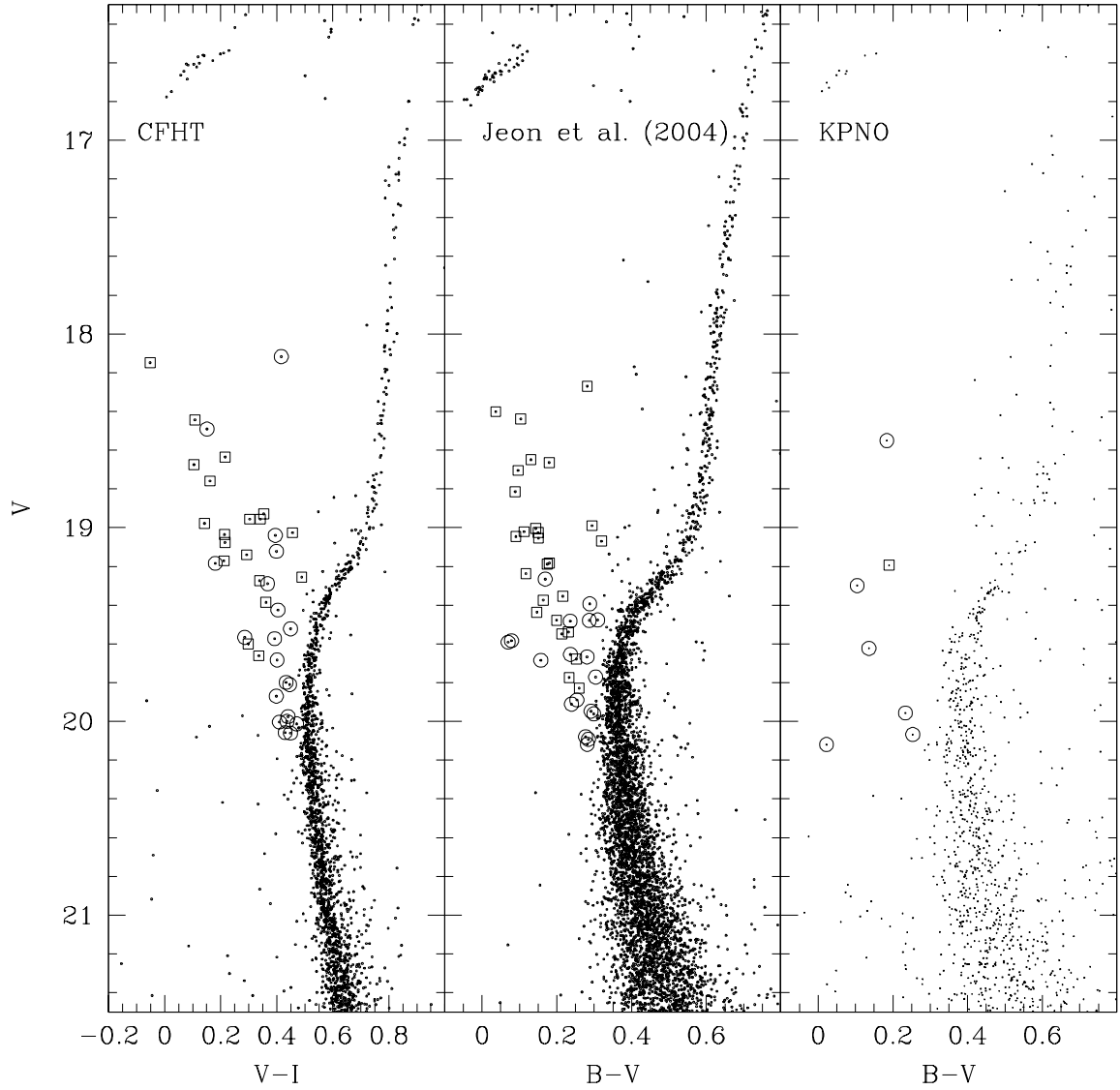}
\caption{Blue straggler selection for NGC 5466. The stars plotted in 
  each panel show the entire sample used for the selection: stars from
  \citet{jeo04} in the middle panel are only those stars outside the CFHT
  field, and KPNO stars in the right panel are only those outside the
  \citeauthor{jeo04} field. Open squares show stragglers identified by
  \citet{nem87}, and open circles are new candidates.
\label{fig555}}
\end{figure}

\clearpage
\begin{figure}
\plotone{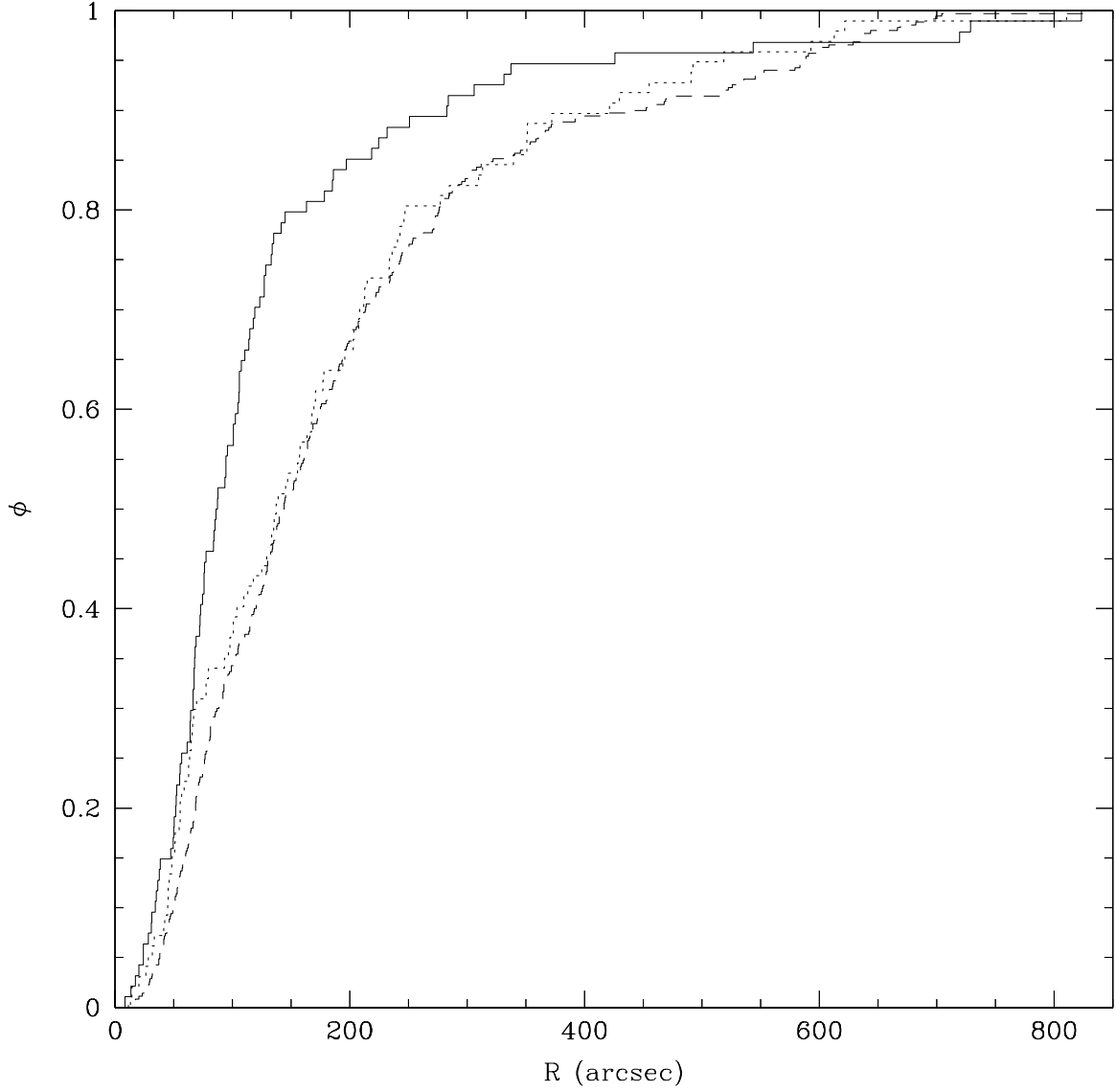}
\caption{Normalized cumulative radial distributions for RGB stars ({\it dashed
    line}), HB stars ({\it dotted line}), and BSSs ({\it solid line}).
\label{fig511}}
\end{figure}

\clearpage
\begin{figure}
\plotone{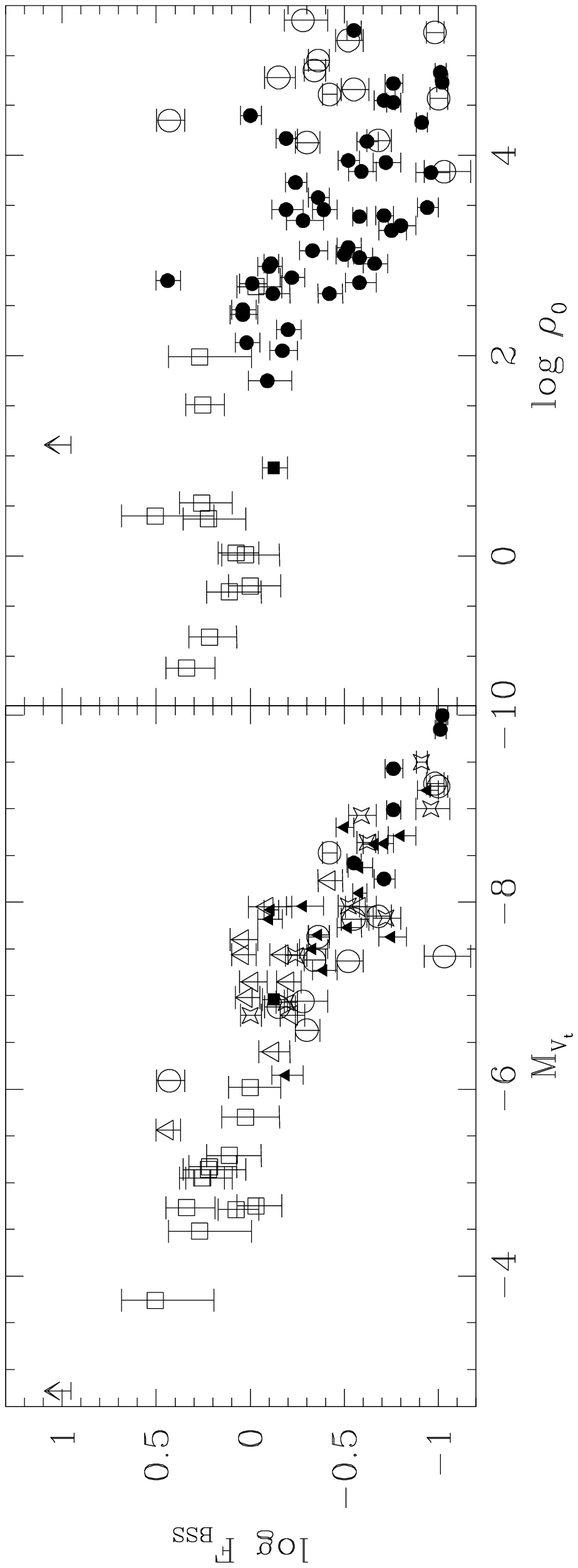}
\caption{Relative frequencies of blue stragglers as a function of
  cluster absolute magnitude and central density. The solid square is NGC
  5466, the open squares are globular clusters from \citet{san05}, and all
  other points are from \citet{pio04}. Open circles are post-core-collapse
  clusters. In the left panel, symbols represent clusters in different ranges
  of central density from the Piotto et al. sample: $\log \rho_0 < 2.8$: open
  triangles; $2.8 < \log \rho_0 < 3.6$: filled triangles; $3.6 < \log \rho_0 <
  4.4$: stars; $\log \rho_0 > 4.4$: filled circles.
\label{fbss}}
\end{figure}

\clearpage
\begin{figure}
\plotone{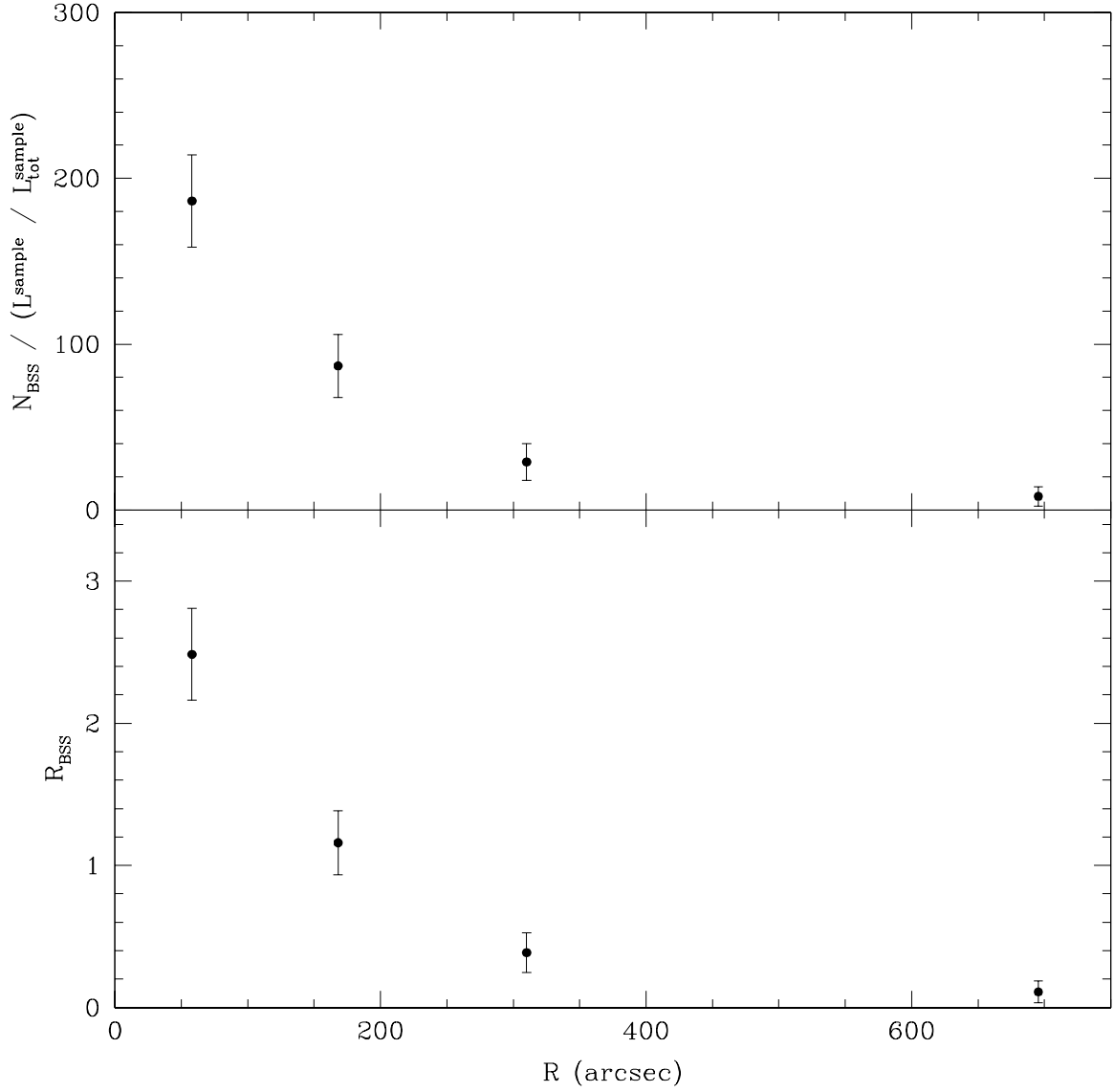}
\caption{Frequency of BSS relative to the integrated $V$-band flux of
detected cluster stars (top panel) and specific frequency of blue stragglers
relative as a function of radius (bottom panel).
\label{fig512}}
\end{figure}


\clearpage
\begin{deluxetable}{rrrrr}
\tablewidth{0pt}
\tablecaption{Photometric Observation Log for NGC 5466 \label{tab31}}
\tablehead{\colhead{UT Date} & \colhead{Filters} & \colhead{$N$} 
& \colhead{Exposure Time (s)} & \colhead{Airmass}}
\startdata
1995 May 4 & $B$,$V$ & 1,1 & 60  & 1.01,1.12 \\
1995 May 4 & $B$     & 2   & 300 & 1.03, 1.03 \\
1995 May 4 & $B$,$V$ & 2,1 & 600 & 1.0,1.11,1.0 \\
1995 May 5 & $B$,$I$ & 2,2 & 300 & 1.03,1.01,1.02,1.01 \\
1995 May 9 & $B$,$V$,$I$ & 1,1,1 & 60 & 1.12,1.13,1.14 \\ 
\enddata
\end{deluxetable} 

\begin{deluxetable}{rrrr}
\tablewidth{0pt}
\tablecaption{$V$-Band Luminosity Function\label{tab51}}
\tablehead{\colhead{$V$} & \colhead{$\log N$} & \colhead{$\sigma_{high}$}
& \colhead{$\sigma_{low}$}}
\startdata
13.532 & 0.5005 & 0.1761 & 0.3010 \\
13.832 & $-0.1015$ & 0.3010 & 1.0000 \\
14.132 & $-0.1015$ & 0.3010 & 1.0000 \\
14.432 & 0.5975 & 0.1605 & 0.2575 \\
14.732 & 0.8985 & 0.1193 & 0.1651 \\
15.032 & 0.6767 & 0.1487 & 0.2279 \\
15.332 & 0.6767 & 0.1487 & 0.2279 \\
15.632 & 0.5976 & 0.1606 & 0.2575 \\
15.932 & 0.8986 & 0.1193 & 0.1651 \\
16.232 & 1.2410 & 0.0839 & 0.1041 \\
16.532 & 1.0127 & 0.1063 & 0.1411 \\
16.832 & 1.0451 & 0.1029 & 0.1351 \\
17.132 & 1.3306 & 0.0764 & 0.0928 \\
17.432 & 1.4432 & 0.0678 & 0.0804 \\
17.732 & 1.4777 & 0.0728 & 0.0876 \\
18.032 & 1.6136 & 0.0630 & 0.0738 \\
18.332 & 1.7575 & 0.0540 & 0.0617 \\
18.632 & 1.7577 & 0.0540 & 0.0617 \\
18.932 & 1.9624 & 0.0433 & 0.0481 \\
19.232 & 2.2928 & 0.0301 & 0.0324 \\
19.532 & 2.5143 & 0.0236 & 0.0249 \\
19.832 & 2.6996 & 0.0192 & 0.0201 \\
20.132 & 2.7738 & 0.0178 & 0.0185 \\
20.432 & 2.8830 & 0.0159 & 0.0165 \\
20.732 & 2.9327 & 0.0151 & 0.0157 \\
21.032 & 3.0036 & 0.0142 & 0.0146 \\
21.332 & 3.0496 & 0.0137 & 0.0142 \\
21.632 & 3.1179 & 0.0132 & 0.0136 \\
\enddata
\end{deluxetable} 

\begin{deluxetable}{rrr}
\tablewidth{0pt}
\tablecaption{Fiducial sequence for NGC 5466 \label{tab34}}
\tablehead{\colhead{$V$} & \colhead{$B-V$} & \colhead{$N$\tablenotemark{a}}}
\startdata
22.198 &       0.578  &       518     \\
21.999 &       0.564  &       598     \\
21.798 &       0.549  &       724     \\
21.596 &       0.532  &       710     \\
21.398 &       0.498  &       673     \\
21.200 &       0.470  &       670     \\
21.004 &       0.459  &       653     \\
20.802 &       0.450  &       614     \\
20.600 &       0.431  &       568     \\
20.400 &       0.418  &       546     \\
20.199 &       0.404  &       470     \\
20.001 &       0.395  &       453     \\
19.805 &       0.396  &       382     \\
19.599 &       0.403  &       273     \\
19.400 &       0.436  &       235     \\
19.206 &       0.495  &       155     \\
19.005 &       0.543  &       84      \\
18.823 &       0.586  &       70      \\
18.588 &       0.607  &       43      \\
18.401 &       0.616  &       52      \\
18.204 &       0.621  &       40      \\
17.989 &       0.636  &       34      \\
17.820 &       0.647  &       31      \\
17.586 &       0.654  &       23      \\
17.403 &       0.672  &       20      \\
17.194 &       0.677  &       19      \\
16.988 &       0.693  &       11      \\
16.784 &       0.722  &       10      \\
16.613 &       0.725  &       6       \\
16.400 &       0.752  &       12      \\
16.199 &       0.771  &       16      \\
16.066 &       0.779  &       5       \\
15.825 &       0.809  &       7       \\
15.602 &       0.827  &       4       \\
15.407 &       0.857  &       6       \\
15.111 &       0.917  &       1       \\
15.006 &       0.930  &       5       \\
14.786 &       0.946  &       2       \\
14.590 &       0.986  &       11      \\
14.438 &       1.044  &       2       \\
\enddata
\tablenotetext{a}{Number of stars used to determine fiducial point}
\end{deluxetable} 

\begin{deluxetable}{lcclll}
\tablewidth{0pt}
\tablecaption{RGB-MSTO Number Ratios \label{tabrat}}
\tablehead{\colhead{Source\tablenotemark{a}} & 
\colhead{TO Sample} & \colhead{RGB Sample} & \colhead{$Y$} &
\colhead{$N_{RGB}/N_{MSTO}$} & \colhead{[Fe/H]}}
\startdata
NGC 5466 & $19.682 < V < 20.282$ & $16.982 < V < 18.482$ & & $0.162\pm0.013$ & \\
VR & & & 0.235 & 0.132 & $-2.22$ \\
VR & & & 0.235 & 0.130 & $-2.14$ \\
T & & & 0.245 & 0.148 & $-2.22$\\
T & & & 0.245 & 0.146 & $-2.14$\\
YY & & & 0.230 & 0.162 & $-2.22$ \\
YY & & & 0.230 & 0.160 & $-2.14$ \\
M3 & $18.80 < V < 19.40$ & $16.40 < V < 18.00$ & & $0.168\pm0.008$ & \\
VR & & & 0.235 & 0.170 & $-1.66$ \\
VR & & & 0.235 & 0.158 & $-1.34$ \\
T & & & 0.246 & 0.181 & $-1.66$\\
T & & & 0.248 & 0.170 & $-1.34$\\
YY & & & 0.230 & 0.185 & $-1.66$ \\
YY & & & 0.230 & 0.162 & $-1.34$ \\
M5 & $19.13 < B < 19.73$ & $16.33 < B < 17.93$ & & $0.110\pm0.006$ & \\
VR & & & 0.235 & 0.106 & $-1.40$ \\
VR & & & 0.235 & 0.097 & $-1.11$ \\
T & & & 0.248 & 0.114 & $-1.40$\\
T & & & 0.251 & 0.106 & $-1.11$\\
YY & & & 0.230 & 0.119 & $-1.40$ \\
YY & & & 0.230 & 0.105 & $-1.11$ \\
M12 & $18.14 < V < 18.74$ & $15.59 < V < 17.24$ & & $0.158\pm0.011$ & \\
VR & & & 0.235 & 0.144 & $-1.40$ \\
VR & & & 0.235 & 0.118 & $-1.14$ \\
T & & & 0.248 & 0.155 & $-1.40$\\
T & & & 0.251 & 0.144 & $-1.14$\\
YY & & & 0.230 & 0.150 & $-1.40$ \\
YY & & & 0.230 & 0.137 & $-1.14$ \\
M30 & $18.33 < V < 18.93$ & $15.78 < V < 17.43$ & & $0.214\pm0.017$ & \\
VR & & & 0.235 & 0.174 & $-2.13$ \\
VR & & & 0.235 & 0.153 & $-1.91$ \\
T & & & 0.245 & 0.173 & $-2.13$\\
T & & & 0.246 & 0.165 & $-1.91$\\
YY & & & 0.230 & 0.194 & $-2.13$ \\
YY & & & 0.230 & 0.180 & $-1.91$ \\
\enddata
\tablenotetext{a}{VR: Victoria-Regina models (no diffusion); YY: Yonsei-Yale
  models (He diffusion); T: Teramo models (no diffusion). All models are for
  an age of 12 Gyr.}
\end{deluxetable} 

\begin{deluxetable}{rrrrrrrrrrcl}
\tablewidth{0pt}
\tablecaption{Selected Star Populations in NGC 5466 \label{tab41}}
\tablehead{\colhead{ID} & \colhead{$\Delta \alpha (\arcsec)$} & 
\colhead{$\Delta \delta (\arcsec)$} & \colhead{$B$} & \colhead{$\sigma_B$} & 
\colhead{$V$} & \colhead{$\sigma_V$} &  \colhead{$I$} & \colhead{$\sigma_I$} &
\colhead{Alternate ID} & \colhead{Ref.\tablenotemark{a}} & \colhead{Notes}}
\startdata
\multicolumn{12}{c}{Blue Stragglers}\\
 1 & 137.74 &  32.86 & 19.1477 &        & 19.0038 &        &         &        &   605 & J & \\
 1 & 137.74 &  32.86 & 19.1446 & 0.0122 & 19.0182 & 0.0280 & 18.7606 & 0.0400 &   809 & K & \\
 2 &$-12.18$& $-5.60$&         &        & 18.7591 & 0.0116 & 18.5972 & 0.0061 & 10313 & C & \\
 2 &$-12.18$& $-5.60$& 18.8461 & 0.0147 & 18.7209 & 0.0258 & 18.6245 & 0.0504 &  2176 & K & \\
 3 & $-8.68$&  14.99 &         &        & 19.1710 & 0.0100 & 18.9595 & 0.0059 &  9339 & C & SX Phe (3)\\
 3 & $-8.68$&  14.99 & 19.4373 & 0.0155 & 19.2441 & 0.0296 & 19.1073 & 0.0563 &  2129 & K & \\
 4 &$-77.31$&  83.56 & 19.2057 &        & 19.0546 &        &         &        &   646 & J & \\
 4 &$-77.31$&  83.56 & 19.2266 & 0.0123 & 19.0548 & 0.0256 & 18.8465 & 0.0460 &  2880 & K & \\
 5 &$-81.70$&  65.06 & 18.5511 &        & 18.2697 &        &         &        &   389 & J & \\
 5 &$-81.70$&  65.06 & 18.5632 & 0.0121 & 18.2740 & 0.0243 & 17.8309 & 0.0255 &  2895 & K & \\
 7 &$-132.96$& $-3.18$& 18.8010 &        & 18.7046 &        &         &        &   504 & J & \\
 7 &$-132.96$& $-3.18$& 18.8482 & 0.0113 & 18.7379 & 0.0217 & 18.5538 & 0.0386 &  3289 & K & \\
 8 &$-144.85$&   3.07 & 18.8436 &        & 18.6640 &        &         &        &   498 & J & \\
 8 &$-144.85$&   3.07 & 18.8726 & 0.0120 & 18.6345 & 0.0234 & 18.3367 & 0.0298 &  3358 & K & \\
 9 &$-90.90$&  83.38 & 19.1757 &        & 19.0256 &        &         &        &   626 & J & \\
 9 &$-90.90$&  83.38 & 19.1385 & 0.0144 & 18.9711 & 0.0257 & 18.7508 & 0.0387 &  2973 & K & \\
10 &$-44.94$&  96.04 & 19.3627 &        & 19.1882 &        &         &        &   751 & J & \\
10 &$-44.94$&  96.04 & 19.4145 & 0.0139 & 19.2190 & 0.0261 & 18.9077 & 0.0494 &  2520 & K & \\
\enddata
\tablenotetext{a}{Photometry Sources: K: KPNO data from this paper, C: CFHT data from this paper, J: \citet{jeo04}}
\tablecomments{The complete version of this table is in the
electronic edition of the Journal. The printed edition contains only a sample.}
\end{deluxetable} 

\end{document}